\documentclass{aa5}
\usepackage{graphicx}
\usepackage{epsfig}
\usepackage{longtable}
\begin{document}
\title{A statistical study of binary and multiple clusters in the
  LMC\thanks{Table \ref{groupscatalogue} is only available in
  electronic form at the 
  CDS via anonymous ftp to {\tt cdsarc.u-strasbg.fr (130.79.125.5)} or via
  {\tt http://cdsweb.u-strasbg.fr/cgi-bin/qcat?J/A+A/}}}

\author{Andrea Dieball\inst{1} \and Hardo M\"uller\inst{2} \and Eva
  K. Grebel\inst{3}} 

\offprints{Andrea Dieball, \email{adieball@astro.uni-bonn.de}}

\institute{Sternwarte der Universit\"at Bonn, Auf dem H\"ugel 71, 
           53121 Bonn, Germany
           \and Institut f\"ur Photogrammetrie, Universit\"at Bonn, 
           Nu\ss allee 15, 53115 Bonn, Germany 
           \and MPI f\"ur Astronomie, K\"onigstuhl 17, 69117
           Heidelberg, Germany} 

\date{Received 20 March 2002 / Accepted 28 May 2002}

\abstract{Based on the Bica et al.\ (\cite{bica}) catalogue, we studied the
  star cluster system of the LMC and provide a new catalogue of all binary and
  multiple cluster candidates found. As a selection criterion we used a
  maximum separation of $1\farcm4$ corresponding to 20 pc (assuming a distance
  modulus of 18.5 mag). We performed Monte Carlo simulations and
  produced  artificial cluster distributions that we compared with the real
  one in order to check how many of the found cluster pairs and groups can be
  expected statistically due to chance superposition on the plane of the
  sky. We found that, depending on the cluster density, between $56 \%$ (bar
  region) and $12\%$ (outer LMC) of the detected pairs can be explained
  statistically. We studied in detail the properties of the multiple cluster
  candidates. The binary cluster candidates seem to show a tendency to form
  with components of similar size. When possible, we studied the age structure
  of the cluster groups and found that the multiple clusters are predominantly
  young with only a few cluster groups older than 300 Myr. The spatial
  distribution of the cluster pairs and groups coincides with the distribution
  of clusters in general; however, old groups or groups with large internal
  age differences are mainly located in the densely populated bar
  region. Thus, they can easily be explained as chance superpositions. Our
  findings show that a formation scenario through tidal capture is not
  only unlikely due to the low 
  probability of close encounters of star clusters, and thus the even lower
  probability of tidal capture, but the few groups with large internal age
  differences can easily be explained with projection effects. We favour a
  formation scenario as suggested by Fujimoto \& Kumai (\cite{fk}) in
  which the components of a binary cluster formed together and thus
  should be coeval or have small age differences compatible with
  cluster formation time scales.  
}

\authorrunning{A. Dieball et al.}
\titlerunning{A statistical study on binary and multiple clusters in the LMC}

\maketitle

\section{Introduction} \label{intro_montecarlo}

The first systematic work on binary clusters in the Magellanic Clouds started
approximately a decade ago. The first catalogue of binary star clusters in
the LMC was presented by Bhatia \& Hatzidimitriou (\cite{bh}) and
Bhatia et al.\ (\cite{brht}) who surveyed the cluster system
(consisting of 1200 objects known at that time) and listed 69 binary cluster
candidates. Their selection criterion was a maximum 
separation between the components of a proposed pair of approximately 18
pc (assuming a distance modulus of 18.4 mag). Following Page
(\cite{page}), out of these 69 double objects only 31 can be explained
as optical pairs, i.e., clusters that appear as close pairs on
the plane of the sky due to projection effects. Ages were available
only for some of the clusters and suggested that the pairs are young
(between $10^{7}$ to a few $10^{8}$ yr), consistent with expected time
scales for merger or disruption of binary clusters (Bhatia \cite{bhatia}). 

In the following years, more studies on binary cluster candidates were
performed but concentrated mainly on one or a few individual objects
in order to establish their binarity (Kontizas et al.\ \cite{kkx}, Lee
\cite{lee92}, Bhatia \cite{bhatia92}, Kontizas et al.\ \cite{kkm},
Vallenari et al.\ \cite{vafcom}, Hilker et al.\ \cite{hrs}, Grebel
\cite{gr97}, Vallenari et al.\ \cite{vbc}, Leon et al.\ \cite{lbv},
Dieball \& Grebel \cite{dg, dg2000}, Dieball et al.\ \cite{dgt}). Few
theoretical studies concerning formation, gravitational interaction
and dynamical evolution of binary clusters are available (Sugimoto \& Makino
\cite{sm}, Bhatia \cite{bhatia}, Fujimoto \& Kumai \cite{fk},
de\,Oliveira et al.\ \cite{dodb}, Theis \cite{theis}).  

However, since the investigation of Bhatia \& Hatzidimitriou (\cite{bh}) and
Bhatia et al.\ (\cite{brht}) many more clusters have been discovered in the
LMC. Thus, it is time to perform a new study on the nowadays better
known LMC cluster system, aiming at the question of how many close
cluster pairs exist and how many of these might be explained
as chance superpositions. 

Recently, Pietrzy\'{n}ski \& Udalski (\cite{pu}) provided a new but
spatially limited catalogue of multiple cluster candidates in the
LMC. They based their studies on the OGLE (see Udalski et al.\
\cite{ogle}) dataset which covers 5.8
square degrees of the inner part of the LMC and contains 745 star
clusters (Pietrzy\'{n}ski et al.\ \cite{puk}). Out of 
these, a total of 100 multiple cluster candidates with a maximum separation of
18 pc, assuming a distance modulus of 18.24 mag, were selected. The cluster
groups consisted of 73 pairs, 18 triple systems, 5 systems containing four
components, 1 with five and 3 systems with six clusters. Assuming that all 745
clusters are distributed uniformly in the 5.8 square degree region and
adopting 
the same statistical approach as Bhatia \& Hatzidimitriou (\cite{bh}), 51
chance pairs can be expected. A more detailed investigation of the cluster
distribution led to nearly the same result of 53 random pairs. The number of
all detected candidates is 153 and thus significantly larger than
expected from chance superposition. Ages for the
components were taken from Pietrzy\'{n}ski \& Udalski (\cite{pu_lmc}). 102
components are coeval, 53 have very different ages, and most objects are
younger than 300 Myr with a peak at 100 Myr. This suggests that most of
the multiple clusters have a common origin and are quite young objects.  

A catalogue of multiple cluster candidates in the SMC was published by
Pietrzy\'{n}ski \& Udalski (\cite{pu_smc}), containing 23 binary and 4
triple cluster candidates. A comparison of both the LMC and SMC binary
cluster lists reveals that the distribution of the components'
separation, the fraction of cluster groups ($\approx$ 12 \%) and their
ages are very similar. The similar ages of the binary cluster
candidates in both the LMC and SMC might be connected with the last
close encounter between these two galaxies. De\,Oliveira et al.\ 
(\cite{odbd}) presented an isophotal atlas of 75 binary and multiple
clusters (comprising 176 objects) from the Bica \&
Dutra (\cite{bd}) catalogue of SMC clusters. Bica \& Dutra (\cite{bd})
included also new discoveries from the OGLE catalogue of SMC clusters
(Pietrzy\'{n}ski et al.\ \cite{puk_smc}). 
Investigating the isophotes of the binary and multiple cluster
candidates, de\,Oliveira et al.\ (\cite{odbd}) found isophotal distortions,
connecting bridges, or common isophotal envelopes for 25 \% of the suggested
multiple clusters. The authors interpreted this as signs of
interaction between the components of a supposed binary or multiple
cluster, in agreement with the findings from previous N-body
simulations (de\,Oliveira et al.\ \cite{obd}).
Ages for 91 out of the 176 clusters that are part of pairs or
groups were investigated based on the OGLE $BVI$ maps. 40 clusters are
in common with Pietrzy\'{n}ski \& Udalski (\cite{pu_smc}), and
de\,Oliveira et al.\ (\cite{odbd}) found good agreement with the
study of the OGLE group. Most clusters are young, and the age
distribution shows a relevant peak around  200 Myr that can be 
attributed to the last close encounter between SMC and LMC. 
The components of groups with more than two members are younger than
100 Myr, which might be an indication that multiple clusters coalesce
into binary or single clusters within this timescale. 55 \% of the binary and
multiple cluster candidates were found to be coeval. From this the authors
concluded that tidal capture is a rare phenomenon.

In this paper, we present a statistical study of close pairs and
multiple clusters in the LMC. We decided to base our analysis on the
new, extended catalogue of stellar clusters, associations, and
emission nebulae in the LMC provided by Bica et al.\ (\cite{bica},
hereafter BSDO). The authors surveyed the ESO/SERC R and J Sky 
Survey Schmidt films, checked the entries of previous catalogues and
searched for new objects. The resolution of the measurements was $<
4\arcsec$ (Bica \& Schmitt \cite{bs}). The resulting new catalogue
unifies previous surveys and contains 6659 entries, out of which
3246 are new discoveries that are not mentioned in previous catalogues and
lists. Thus, the BSDO catalogue can be considered as the
so far most complete catalogue of LMC stellar clusters and associations. 
We restricted our study to bound stellar systems, which
means that we selected only objects which are categorized as ``C''-type
(cluster-type), and left out associations and emission nebulae, which are not
of interest in the context of the present study. This reduces the number of
objects found in the BSDO catalogue from a total of 6659
to 4089. 

Based on this catalogue, we performed a statistical study of cluster pairs and
groups and provide a complete list of all multiple cluster candidates in the
LMC.   

In the following sections, we address a number of questions:
How many cluster pairs can be found with a projected separation of
less than 20 pc between the components of a pair
(Sect.~\ref{groups_montecarlo})? Following Bhatia \& 
Hatzidimitriou (\cite{bh}) and Bhatia et al.\ (\cite{brht}), we
consider this to be a good selection criterion. Several cluster pairs
may form a larger cluster group, e.g., if a component of a cluster
pair is less than 20 pc distant from any component of another pair. In
this way, three clusters may form a triple cluster, but they also
might constitute three cluster pairs if each cluster is seen within 20
pc from each other cluster. How many ``multiple'' clusters, consisting
out of more than two single objects, are present, and how many single
clusters are involved in these pairs and groups
(Sect.~\ref{groups_montecarlo})? How many of these pairs and multiple  
systems can be expected statistically, and of how many individual
components do they consist (Sect.~\ref{sim_montecarlo})? Are there any
correlations between the properties of the cluster systems such as 
ages, radii and separations between the components
(Sect.~\ref{groupsproperties})? What is the fraction of coeval pairs
or groups compared with the number of multiple clusters whose 
internal age differences exceed the protocluster survival time
(Sect.~\ref{pairsages})? Does the percentage of coeval systems agree
with the number of statistically expected groups
(Sect.~\ref{pairsages})? And finally, do our results favour or 
give hints at a specific cluster formation scenario
(Sect.~\ref{summary_montecarlo})? For instance, can cluster pairs be
explained with statistically expected cluster encounters in the LMC,
which lead to tidal capture and thus to bound pairs of different ages
(see Sect.~\ref{encounter_montecarlo})? Or are the multiple cluster
candidates predominantly found to be coeval, favouring the formation scenario
of Fujimoto \& Kumai (\cite{fk}) or of Theis (\cite{theis})?  

\section{Cluster density distribution} \label{cldensity_montecarlo}

Looking at an optical image of the LMC, it
is apparent that the stellar and cluster distribution is not uniform across
this galaxy. One of the most striking features besides the prominent
star-forming \ion{H}{ii} regions is the LMC bar, a region of increased stellar
density. 

\begin{figure}[]
\centerline{
\epsfig{file=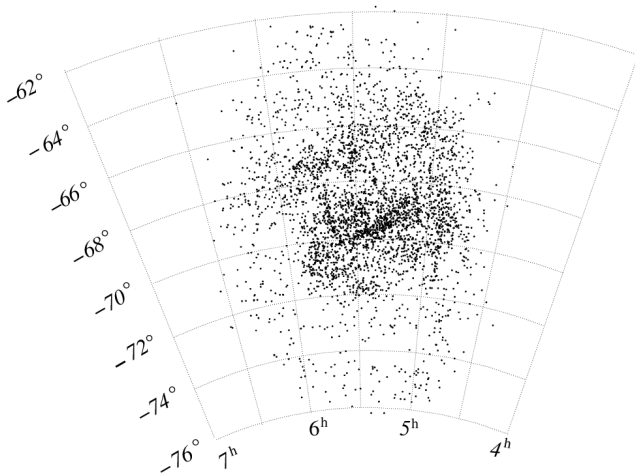, width=8.5cm, clip=}
}
\caption[Angular distribution of LMC
  clusters]{\label{bicaclusters} Angular 
  distribution of all cluster-like objects derived from the BSDO
  catalogue. The bar structure is clearly visible}   
\end{figure}

\begin{figure}[]
\centerline{
\epsfig{file=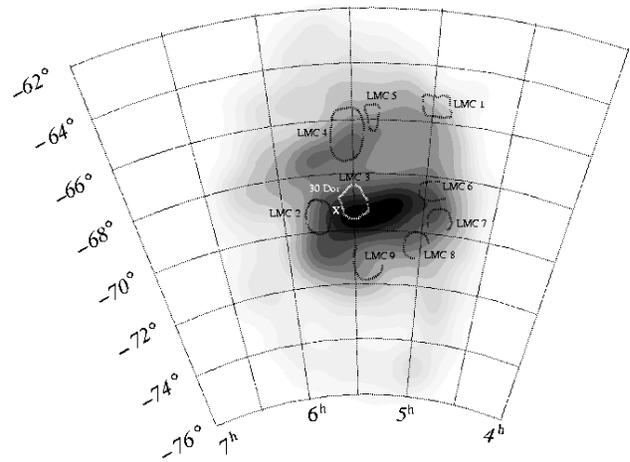, width=8.5cm, clip=}
}
\caption[Cluster density plot for the
  LMC]{\label{bicaclustersdens} Density 
  plot of all LMC clusters. The highest cluster density (black) coincides with
  the bar region. Please note that this picture provides qualitative
  information about the density distribution. No quantitative values of the
  cluster densities can be derived from the image's greyscales alone. The
  location of the supergiant shells LMC\,1 -- LMC\,9 are sketched (see Fig.~6,
  page 6, in Braun \cite{braundr}). The position of 30\,Doradus is marked with
  a white cross}
\end{figure}

In Fig.~\ref{bicaclusters} we plotted the angular distribution of all
clusters listed in the BSDO catalogue. Again, the bar
structure can easily be recognized. However, the clusters are not evenly
distributed in the outer LMC. To make the structures of the cluster
density distribution more apparent, Fig.~\ref{bicaclusters} was smoothed
with a Gaussian filter with a blur radius of 50 image pixels. The
intensity scale of the resulting figure (Fig.~\ref{bicaclustersdens})
was reduced to 15 bins, the darkest shade indicating the highest
cluster density. In this way, it is possible to get qualitative
information about the density distribution, however, quantitative values of the
cluster densities cannot be derived from the image's greyscales alone. 
All steps were performed with the aid of common image processing tools (GIMP). 

Regions of different cluster densities and their spatial extent can be seen in
Fig.~\ref{bicaclustersdens}: Apart from the prominent bar structure, also the
area surrounding the bar is densely populated with star clusters. To the
north-east another region of enhanced cluster density is clearly visible,
which coincides with LH\,77 (Lucke \& Hodge \cite{lh}) in the supergiant shell
LMC\,4 (see e.g. Braun et al.\ \cite{braun}) and the constellation Shapley\,III
(McKibben Nail \& Shapley \cite{mshapley}). The outermost LMC
areas show a considerable drop-off in the cluster density, which is
already recognizable in Fig.~\ref{bicaclusters}. 

\begin{figure}[]
\centerline{
\epsfig{file=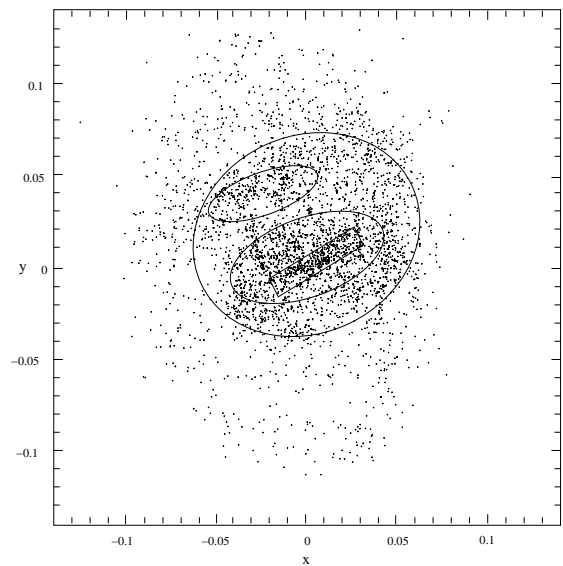, width=8.5cm, clip=}
}
\vspace{-0.5cm}
\caption[LMC cluster distribution in cartesian
coordinates]{\label{bicaclustersxy} Cluster distribution plotted in cartesian
  coordinates. The regions that are selected on the basis of the cluster
  density plot are sketched}   
\end{figure}

Due to the non-uniform distribution of the star clusters in the LMC, we
decided to subdivide the BSDO catalogue into
regions of equal cluster densities. It is a difficult task to find a
suitable partition: it has to be small enough so that a (nearly)
uniform distribution of clusters can be assumed but general density
differences between larger areas must still be recognizable and may
not be smeared out by a too small partition, and the
regions have to be larger than the detection limit for cluster pairs
and groups. Based on Fig.~\ref{bicaclustersdens}, we decided to split the
input catalogue into five lists: the inner part of the LMC shows in general a
higher concentration of star clusters, thus we first distinguish between an
inner ellipse (which we call $E_{\rm inner}$) and an outer ring, the
remaining outer LMC area which shows an overall low cluster density ($E_{\rm
  outer}$). Inside the inner ellipse, there are still regions of varying 
cluster frequencies. Thus, we further define an ellipse surrounding the bar
(which we call $E_{\rm bar}$), a rectangular area that coincides with the
bar itself (called ``bar''), and an ellipse north-east of the bar
corresponding to the location of LMC\,4 ($E_{\rm north}$). All areas are
disjunctive, i.e., if a selected region contains one or more selected ellipses
or the bar, these ``inner'' areas are not considered in the following
calculations and simulations, i.e., $E_{\rm bar}$ does not contain the bar,
$E_{\rm inner}$ does not contain $E_{\rm bar}$ and $E_{\rm north}$ and
so on.  

For the following selections and Monte Carlo simulations it is suitable to
transform the spherical coordinates of the catalogue entries ($\alpha$,
$\delta$ coordinates) into cartesian coordinates ($x, y$) (see, e.g.,
Geffert et al.\ \cite{geffert} and Sanner et al.\ \cite{sanner}). 
The gnomonic projection was done via:
\begin{eqnarray}
x &=& \frac{-\cos\delta \cdot \sin(15 \cdot (\alpha-\alpha_{0}))}{\sin\delta \cdot \sin\delta_{0}+\cos\delta \cdot \cos\delta_{0} \cdot \cos(15 \cdot (\alpha-\alpha_{0}))} \label{transx}\\\nonumber\\
y &=& \frac{\sin\delta \cdot \cos\delta_{0}-\cos\delta \cdot \sin\delta_{0} \cdot \cos(15 \cdot (\alpha-\alpha_{0}))}{\sin\delta \cdot \sin\delta_{0}+\cos\delta \cdot \cos\delta_{0} \cdot \cos(15 \cdot (\alpha-\alpha_{0}))} \label{transy}
\end{eqnarray}
(van\,de\,Kamp \cite{vandekamp}).
We adopt $\alpha_{2000} = -69^{\circ}$ $45\arcmin$ and $\delta_{2000} = 5^{h}$
$23\farcm6$ as the central coordinates of the LMC (CDS data
archive). Fig.~\ref{bicaclustersxy} 
shows the cluster distribution in cartesian coordinates.  

The selection criterion for clusters situated inside an ellipse with the semi
major axes $a$ and $b$, central coordinates $x_{0}$ and $y_{0}$, and rotation
angle $\phi$ is: 
\begin{eqnarray}
\label{ellips}
&\left(\frac{(x-x_{0})\cdot\cos\phi+(y-y_{0})\cdot\sin\phi)}{a}\right)^{2}+&\\\nonumber
&\left(\frac{(x-x_{0})\cdot\cos\phi-(y-y_{0})\cdot\sin\phi}{b}\right)^{2}\leq&1 \end{eqnarray}
The semi axes, central coordinates and the rotation angle for the selected
ellipses are listed in Table~\ref{ellipscoord}. A helpful tool for selecting
the bar region is the map of the LMC provided by Smith et al.\ (\cite{smith},
their Fig.~4) in which prominent features are sketched. The vertices that we
choose to cut out the bar are listed in Table~\ref{barcoord}.

All selected areas are sketched in Fig.~\ref{bicaclustersxy}.

\begin{table}[]
\caption[Parameters defining the selected LMC
regions]{\label{ellipscoord} Semi axes $a$ and $b$ (in units of the
cartesian system), central coordinates $x_{0}$ and $y_{0}$ and the
rotation angle $\phi$ for the selected ellipses for which uniform
cluster densities are assumed (see Eq.~\ref{ellips})} 
\centerline{
\begin{tabular}{lccrcc}
\hline
Region            & $a$   & $b$   & \multicolumn{1}{c}{\mbox{ } $x_{0}$} & $y_{0}$ & $\phi$       \\
\hline
$E_{\rm bar}$   & 0.045 & 0.020 & 0.000    & 0.005   & $20^{\circ}$ \\
$E_{\rm north}$ & 0.030 & 0.013 & 0.025    & 0.040   & $20^{\circ}$ \\
$E_{\rm inner}$ & 0.065 & 0.055 & 0.002    & 0.018   & $25^{\circ}$ \\
$E_{\rm outer}$ & 0.100 & 0.135 & $-0.005$ & 0.015   & $0^{\circ}$ \\
\hline
\end{tabular}
}
\end{table}  

\begin{table}[]
\caption[Vertices defining the LMC bar region]{\label{barcoord} Vertices
  defining the selected bar region} 
\centerline{
\begin{tabular}{rr}
\hline
\multicolumn{1}{c}{$x$} & \multicolumn{1}{c}{$y$}      \\
\hline
0.0265    & 0.0212 \\
0.0337    & 0.0119 \\
$-0.0199$ & $-0.0067$\\
\hline
\end{tabular}
}
\end{table}  

\section{Probability of tidal capture} \label{encounter_montecarlo}

The probability of close encounters between star clusters leading to
a tidal capture is considered to be relatively small or even very unlikely
(Bhatia et al.\ \cite{brht}). However, van den Bergh (\cite{vdbergh})
suggested that it becomes more probable in dwarf galaxies like the
Magellanic Clouds due to the small velocity dispersion of the cluster
systems. Furthermore, Vallenari et al.\ (\cite{vbc}) proposed that
interactions between the LMC and SMC might have increased the
formation of star clusters in large groups in which the encounter rate
and thus the formation of bound binary clusters is higher. This
scenario is capable of explaining large age differences between
cluster pairs which Leon et al.\ (\cite{lbv}) refer to as the
``overmerging problem''. In this section, we will determine the
cluster encounter rates in our selected areas. 

Inside the chosen regions we find 491 clusters in the bar, 863 in the
remaining parts of $E_{\rm bar}$, 372 objects in $E_{\rm north}$, 1439
clusters in $E_{\rm inner}$ (without $E_{\rm bar}$ and $E_{\rm north}$),
and 924 remaining entries in the outer region of the LMC. 
Assuming a distance modulus of 18.5 mag to the LMC, 1~pc corresponds to
$1.999\cdot10^{-5}$ units in our cartesian system. This leads to a length of  
2710 pc and a width of 591 pc for the bar, and thus to a cluster density of
$3\cdot10^{-4}$ clusters${\rm ~pc}^{-2}$. For $E_{\rm bar}$, the semi major
axis corresponds to 2250 pc and its semi minor axis to 1000 pc, resulting in 
$863 \mbox{ }{\rm clusters}/(A_{\rm E_{\it bar}}-A_{\rm bar}) =
1.6\cdot10^{-4}$ clusters${\rm ~pc}^{-2}$. Cluster densities for the other
selected areas follow in an analogous manner: $1.2\cdot10^{-4}$ clusters${\rm ~pc}^{-2}$ for $E_{\rm north}$, $8\cdot10^{-5}$ clusters${\rm ~pc}^{-2}$
for $E_{\rm inner}$ (without the northern ellipse and the region surrounding
the bar), and $1.2\cdot10^{-5}$ clusters${\rm ~pc}^{-2}$ for the outer ring
$E_{\rm outer}$. Please note that, assuming an outer {\it ring} with
limited boundaries and semi axes corresponding to 5000 pc and 6750 pc, the
number of objects in $E_{\rm outer}$ amounts to 911. However, this does not
alter the value of the outer cluster density. 

The cluster density is highest in the innermost part of the LMC, the bar
region, and it drops off by an order of magnitude towards the outer
region. According to Vallenari et al.\ (\cite{vbc}), cluster pairs can be
formed by close encounters which result in the tidal capture of two
clusters. The higher the cluster density, the higher the probability for
close encounters, and thus the probability for the formation of cluster pairs
or groups. 

The cluster encounter rate can be determined following Lee et
al.\ (\cite{lee95}):  
\begin{eqnarray}
&&\frac{{\rm d}N}{{\rm d}t} = \frac{1}{2} \cdot \frac{N-1}{V} \cdot \sigma \cdot v 
\end{eqnarray}
where $N$ is the number of clusters, $V$ denotes the volume of the galaxy or,
respectively, the part of the galaxy under investigation, $\sigma = \pi R^{2}$
is the geometric cross section of a cluster with radius $R$, and $v$ is the
velocity dispersion of the cluster system of that galaxy. 
Typical cluster radii are about 10 pc. For the velocity dispersion of the
cluster system we adopt 15 km $\mbox{s}^{-1}$ as quoted in Vallenari et
al.\ (\cite{vbc}). For the depth of the LMC, and thus for each selected area,
we adopt 400 pc (Hughes et al.\ \cite{hughes}). This leads to a cluster
encounter 
rate of $20\cdot 10^{-10}~{\rm yr}^{-1}$ inside the bar. In the ellipse
surrounding the bar, the northern region and the inner LMC region, the
probability of close encounters is much lower by a factor of 2 to 4, namely 
$9\cdot 10^{-10}~{\rm yr}^{-1}$  in $E_{\rm bar}$, $7\cdot
10^{-10}~{\rm yr}^{-1}$ in $E_{\rm north}$, and $5\cdot 10^{-10}~{\rm
  yr}^{-1}$ in $E_{\rm inner}$. The lowest encounter rate is, as
expected, in the outer ring with a value of $0.7\cdot 10^{-10}~{\rm 
  yr}^{-1}$. This means that the probability of a close encounter
between star clusters is $\approx 30$ times higher in the bar than in the
outskirts of the LMC. 

All results are summarized in Table~\ref{encounter}.

The probabilities for cluster encounters are already very low. In addition, the
probability of {\it tidal capture} depends on further conditions which will
not be fulfilled during every encounter. Whether a tidal capture takes place
or not depends strongly on the velocities of the two clusters with respect
to each other, on the angle of incidence, whether sufficient angular
momentum can be transferred, and whether the clusters are sufficiently
massive to survive the encounter. Since only very few of these rare
encounters would result in tidal capture, it seems unlikely that a
significant number of young pairs may have formed in such a scenario. 
      
\begin{table*}[ht!]
\caption[Cluster densities and encounter rate in the selected
regions]{\label{encounter} Semi axes $a$, $b$ in pc and number of clusters
  found in the selected regions, the resulting cluster densities and the
  encounter rate ${\rm d}N/{\rm d}t$. For the bar region, $a$
  and $b$ do not denote semi axes, but 
  the lengths of a rectangular area}  
\centerline{
\begin{tabular}{lclrrr}
\hline
Region          & \multicolumn{1}{c}{$a$ [pc]} & \multicolumn{1}{c}{$b$ [pc]} & \multicolumn{1}{c}{$N_{\rm clusters}$} & \multicolumn{1}{c}{$\frac{{\rm clusters}}{{\rm ~pc}^{2}}$} & \multicolumn{1}{c}{${\rm d}N/{\rm d}t$} \\
\hline
Bar             & (2710) & (591)                  & 491  & $3\cdot10^{-4}$   & $20\cdot 10^{-10}{\rm yr}^{-1}$\\
$E_{\rm bar}$   & 2250   & 1000                   & 863  & $1.6\cdot10^{-4}$ & $9\cdot 10^{-10}{\rm yr}^{-1}$\\
$E_{\rm north}$ & 1500   & $\mbox{ }\mbox{\,}650$ & 372  & $1.2\cdot10^{-4}$ & $7\cdot 10^{-10}{\rm yr}^{-1}$\\
$E_{\rm inner}$ & 3250   & 2750                   & 1439 & $8\cdot10^{-5}$   & $5\cdot 10^{-10}{\rm yr}^{-1}$\\
$E_{\rm outer}$ & 5000   & 6750                   & 911  & $1.2\cdot10^{-5}$ & $0.7\cdot 10^{-10}{\rm yr}^{-1}$\\
\hline
\end{tabular}
}
\end{table*} 

\section{Numbers of cluster pairs and cluster groups}
\label{groups_montecarlo}

\begin{table*}[ht!]
\caption[Cluster group statistics in the selected regions]{\label{groupstat}
  Statistics about the cluster groups found in the 
  selected regions. $N_{\rm tot}$ denotes all clusters found in the
  corresponding region, $N_{\rm cl}$ is the number of clusters involved
  in $N_{\rm pairs}$ pairs. The numbers of isolated pairs $N_{\rm 2}$, triple
  systems $N_{\rm 3}$, and so on can be found in the subsequent
columns. Groups consisting of more than eight members do not occur}  
\centerline{
\begin{tabular}{lrrrrrrrrrr}
\hline
Region            & \multicolumn{1}{c}{$N_{\rm tot}$} &
\multicolumn{1}{c}{$N_{\rm cl}$} & \multicolumn{1}{c}{$N_{\rm pairs}$} &
\multicolumn{1}{c}{$N_{\rm 2}$} & \multicolumn{1}{c}{$N_{\rm 3}$} &
\multicolumn{1}{c}{$N_{\rm 4}$} & \multicolumn{1}{c}{$N_{\rm 5}$} &
\multicolumn{1}{c}{$N_{\rm 6}$} & \multicolumn{1}{c}{$N_{\rm 7}$} &
\multicolumn{1}{c}{$N_{\rm 8}$}  \\ 
\hline
Bar               &  491 &  228 & 166 &  59 & 22 &  5 & -- &  4 & -- & -- \\
$E_{\rm bar}$   &  863 &  306 & 207 &  97 & 20 &  5 &  2 &  1 & -- & 2  \\
$E_{\rm north}$ &  372 &  117 &  88 &  36 &  5 &  3 &  2 & -- & -- & 1  \\
$E_{\rm inner}$ & 1439 &  371 & 247 & 131 & 19 &  5 &  4 &  2 & -- & -- \\
$E_{\rm outer}$ &  924 &   93 &  55 &  40 &  3 &  1 & -- & -- & -- & -- \\
\hline
Sum               & 4089 & 1115 & 763 & 363 & 69 & 19 &  8 &  7 & -- & 3  \\
\hline
LMC total         & 4089 & 1126 & 770 & 366 & 69 & 19 &  9 &  7 & -- & 3  \\
\hline
\end{tabular}
}
\end{table*}  

As the selection criterion for binary cluster candidates we chose a maximum
angular separation of $\le 1\farcm4$ corresponding to a projected distance of
20 pc (assuming a distance modulus of 18.5 mag) between the centres of
a proposed cluster pair. This is nearly the same value that was used
by Bhatia \& 
Hatzidimitriou (\cite{bh}), Hatzidimitriou \& Bhatia (\cite{hb}), and Bhatia
et al.\ (\cite{brht}). According to Bhatia (\cite{bhatia}) and Sugimoto \&
Makino (\cite{sm}), binary clusters with larger separations may become detached
by the external tidal forces while shorter separations may lead to mergers.  

Out of 491 clusters in the bar region, 228 objects can be found in double or
even multiple systems. This amounts to $\approx 46 \%$ of all bar clusters.
We counted in total 166 pairs. However, two or more pairs may
form a larger group, e.g., three star clusters may form
up to three pairs if each cluster is seen within a projected distance of less
than 20 pc from each other cluster. This means that the 166 pairs do not
consist of 334 different single clusters but only of the 228 objects mentioned
above. Hence we call only an isolated pair a possible binary
system. In the bar 59 isolated pairs, 22 triple clusters and 9 larger
groups with up to six members can be found.  

The area surrounding the bar, $E_{\rm bar}$, is roughly half as densely
populated with star clusters as the bar region itself. The percentage of
clusters found in potential binary and multiple systems is still high,
$\approx 35 \%$ (306 objects), forming in total 207 pairs. 
       
The cluster density in the northern region, $E_{\rm north}$, is nearly the
same as in $E_{\rm bar}$, and approximately the same {\bf percentage}
of clusters 
($\approx 31 \%$ or 117 objects) can be found in 88 pairs which form 36
binary, 5 triple, and 6 larger systems.   

In the remaining inner LMC region ($E_{\rm inner}$) the cluster density is
lower by an order of a magnitude; however, still $\approx 26\%$ (371) of the
clusters appear in potential binary and multiple systems. 

In the outskirts of the LMC, the cluster density is the lowest, as is
the number of clusters involved in pairs and groups: 93 ``outer''
clusters, i.e., $10\%$, form in total 55 
pairs (40 binary, 3 triple and 1 quadruple systems).

The distribution of all cluster groups found in each selected area is
summarized in Table~\ref{groupstat} and illustrated in
Fig.~\ref{grouphisto}. The percentage of all clusters involved in the
groups is indicated in Fig.~\ref{grouphisto}, e.g., $24 \%$ (or 118)
of all clusters can be found in 59 binary systems in the bar.  

Table~\ref{groupstat} also lists the sum of all clusters and cluster pairs
which result if the values for the different regions are added up. The last
line of Table~\ref{groupstat} gives the group statistics for the whole LMC
without a division into separated areas. As can be seen, the sum of the
individual group statistics differs from the statistics found for the total
LMC. This is due to the fact that some multiple cluster candidates are located
at or across the borders of the selected areas, so that they get divided into
smaller groups or might even disappear as a cluster group through the division
into different regions.

We caution that this statistical approach so far does not take into account
possible age differences between clusters (which are mainly unknown).  Also,
we do not have any other information about the actual
three-dimensional separation between the clusters.  

\section{Monte Carlo Simulations -- how many pairs can be expected
  statistically?} 
\label{sim_montecarlo}

In order to check how many of all the found pairs and multiple clusters can be
expected statistically due to chance line-up, we performed statistical
experiments:  

For this purpose we developed {\tt C} and {\tt C++} software which performs
Monte Carlo simulations and analyses of the resulting random distribution. 

The simulations are carried out in the cartesian system and we used the same
selected areas as mentioned in Sect.~\ref{cldensity_montecarlo}. The same
number of objects which were found based on the BSDO
catalogue are now distributed randomly in each region, i.e., 372 objects are
randomly spread in an ellipse with semi axes of 0.030 and 0.013 (or 1500 and
650 pc) for the inner, northern region $E_{\rm north}$; 491 objects are
stochastically distributed in a rectangular area with lengths of
$1.1822\cdot10^{-2}$ and $5.4189\cdot10^{-2}$  in units of the cartesian system
(corresponding to 591 and 2710 pc) which denotes the LMC bar; 863 objects are
arbitrarily placed in the space between the bar and the boundaries of the
ellipse described with $E_{\rm bar}$ and so on. In this way, artificial
cluster distributions are produced that can be compared with the true
distribution in the LMC. To improve statistics this procedure was repeated 100
times for each region.     

\begin{figure}[]
\centerline{
\epsfig{file=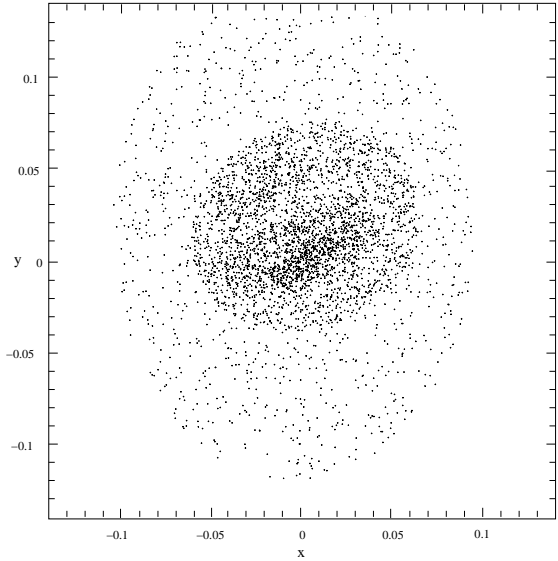, width=8.5cm, clip=}
}
\caption[Artificial cluster distribution]{\label{simxy} Example of one of the
  artificial cluster distributions that was created using Monte Carlo
  simulations. The distribution is plotted in cartesian coordinates and can be
  compared with Fig.~\ref{bicaclustersxy}, which shows the real LMC cluster
  distribution}   
\end{figure}

\begin{figure}[]
\centerline{
\epsfig{file=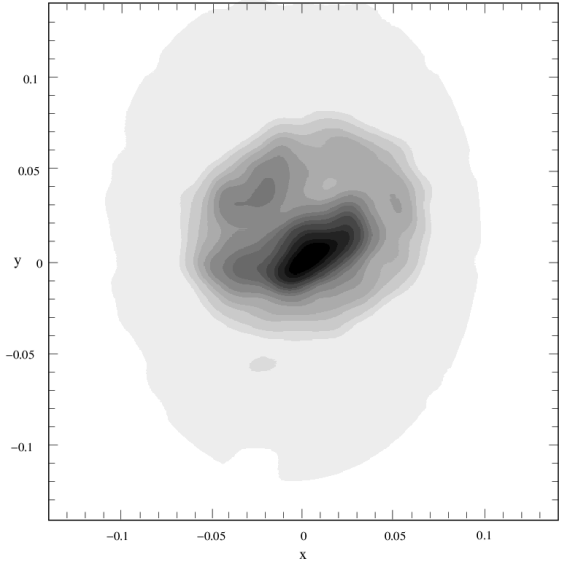, width=8.5cm, clip=}
}
\caption[Density distribution of the artificial cluster
system]{\label{simdensxy} Density distribution of the artificial cluster 
  system represented in Fig.~\ref{simxy}}  
\end{figure}

An example of an artificial cluster distribution is plotted in
Fig.~\ref{simxy}. The inner part of the LMC, including the bar and the
northern region corresponding to LMC\,4, is well represented. However, there
is a sharp drop-off in the cluster density in the outskirts. The cluster
density in the LMC is low in its outer regions, however, the decrease between
the inner and outer region ($E_{\rm inner}$ and $E_{\rm outer}$) is
smoother (see Fig.~\ref{bicaclustersxy}). Fig.~\ref{simdensxy} displays the
density distribution of the artificial cluster system. Compared with
Fig.~\ref{bicaclustersdens} all prominent features are well represented.

The number of chance pairs and groups  in our simulated cluster distributions
as well as the number of single objects involved in these groups was counted
and compared with our findings based on BSDO. As pointed
out in the previous sections, two objects are considered as a pair and thus 
will be included in our list of random pairs or groups if their
separation is 20 pc ($3.9979\cdot10^{-4}$ in units of the cartesian
system) at maximum. 

Out of 491 random objects in the bar region, $\approx 156 \pm 14$ can be found
in $\approx 94 \pm 10$ pairs. In reality, the LMC bar comprises 166 pairs which
is nearly twice the amount of what we can expect statistically. However, a
closer look at the {\it group} statistics reveals that of the 59 binary
systems found in the real LMC bar cluster distribution, $55 \pm 6$ isolated
pairs can be explained as chance pairs based on our simulations, i.e.,
these numbers agree within the uncertainties. A discrepancy between
found and expected 
figures is more apparent when groups with three or more 
members are considered. Out of 22 triple systems, only $\approx 11 \pm 4$
(i.e., $\approx 50 \%$) can be explained with chance superpositions. The number
of actually found groups containing four members is twice as much as expected
(5 found, 2.49 expected). 4 groups with six members are found,
however, their random formation is very unlikely (0.21).   

\begin{figure}
\includegraphics[width=\hsize]{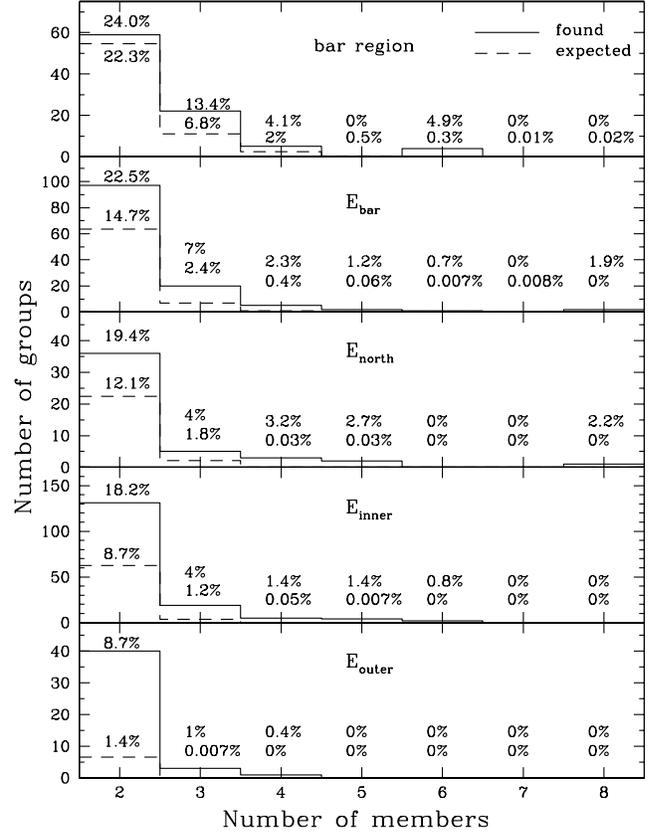}
\caption[Cluster groups in the selected regions]{\label{grouphisto} Histogram
  of the number of cluster groups found in the selected regions of different
  cluster densities. The solid line denotes the number of groups detected in
  the real cluster distribution, while the dashed line indicates the number of
  cluster systems which can be expected statistically. The percentage of the
  clusters involved in the groups of different member size is also given. See
  Sect.~\ref{sim_montecarlo} for details}
\end{figure}

The statistically expected number of chance pairs and groups are summarized in
Table~\ref{simgroupstat} for all regions. A graphical display of our results
can be found in Fig.~\ref{grouphisto}. The numbers of the cluster
groups actually found in the LMC regions are also indicated in
Fig.~\ref{grouphisto} (see also Sect.~\ref{groups_montecarlo}) so that
the results of both the real and the simulated cluster distribution
can easily be compared.  

In the remaining space of $E_{\rm bar}$, 207 pairs are actually found, but
only $84 \pm 9$ ($\approx 41 \%$) of them can be explained statistically. Out
of 97 isolated pairs, $\approx 66 \%$ can be ascribed to chance
superpositions. 20 triple systems are found, but only $7 \pm 2$ can be
expected. Again, for larger groups the discrepancy increases. 

In the northern region, $E_{\rm north}$, $33 \%$ ($\approx 29 \pm 6$ pairs) of
the 88 ones found can be expected due to chance line up.

For the remaining part of the inner LMC ($E_{\rm inner}$), $\approx 29 \%$
of all pairs are explainable on a statistical basis.  

\begin{table*}[ht!]
\caption[Statistics about the cluster groups in the selected
regions]{\label{simgroupstat} Statistics about the cluster groups found in 
  the simulated regions. The standard deviations are given in brackets. Column
  labels are as in Table~\ref{groupstat}}    
\centerline{
\begin{tabular}{lrrrrrrrrrr}
\hline
Region            & \multicolumn{1}{c}{$N_{\rm tot}$} &
\multicolumn{1}{c}{$N_{\rm cl}$} & \multicolumn{1}{c}{$N_{\rm pairs}$} &
\multicolumn{1}{c}{$N_{\rm 2}$} & \multicolumn{1}{c}{$N_{\rm 3}$} &
\multicolumn{1}{c}{$N_{\rm 4}$} & \multicolumn{1}{c}{$N_{\rm 5}$} &
\multicolumn{1}{c}{$N_{\rm 6}$} & \multicolumn{1}{c}{$N_{\rm 7}$} &
\multicolumn{1}{c}{$N_{\rm 8}$}  \\ 
\hline
Bar               &  491 & 156.39 & 94.18 & 54.64 & 11.13 & 2.49 & 0.47 & 0.21
& 0.01 & 0.01 \\
                  &      & (14.28)& (9.83)& (6.11)&(3.67)&(1.72)&(0.67)&(0.46)
&(0.10) & (0.10) \\
$E_{\rm bar}$   &  863 & 152.37 & 84.05 & 63.59 &  7.00 & 0.89 & 0.10 & 0.01
& 0.01 & -- \\
                  &      & (15.57)& (9.23)& (7.20)& (2.40)&(0.84)&(0.30)&(0.01)
&(0.01) & -- \\
$E_{\rm north}$ &  372 &  52.72 & 28.83 & 22.44 &  2.18 & 0.30 & 0.02 & --
& --   & -- \\
                  &      & (10.24)& (5.70)& (4.41)& (1.34)&(0.56)&(0.14)& --
& --   & -- \\
$E_{\rm inner}$ & 1439 & 136.51 & 71.50 & 62.53 &  3.53 & 0.19 & 0.02 & --
& --   & -- \\
                  &      & (16.20)& (8.61)& (7.59)& (1.77)&(0.44)&(0.14)& --
& --   & -- \\
$E_{\rm outer}$ &  911 &  13.12 &  6.58 &  6.53 &  0.02 & --   & --   & --
& --   & -- \\
                  &      & (5.16) & (2.55)& (2.54)& (0.14)& --   & --   & --
& --   & -- \\
\hline
Sum               & 4076 & 511.11 & 285.14 & 209.73 & 23.86 & 3.87 & 0.61 &
0.22 & 0.02 & 0.01 \\
                  &      & (28.82)& (14.92)& (12.12)& (5.10)&(2.16)&(0.85)&(0.48)&(0.14)&(0.20) \\
\hline
Total             & 4076 & 515.58 & 288.01 & 211.32 & 24.19 & 3.97 & 0.63 &
0.22 & 0.02 & 0.02 \\
                  &      & (26.02)& (15.35)& (12.2) & (4.96)& (2.49) & (0.80) &
(0.46)& (0.14) & (0.14) \\
\hline
\end{tabular}
}
\end{table*}  

The outer LMC is approximated by a ring (see Sect.~\ref{encounter_montecarlo})
in which 911 objects are randomly distributed. These are 13 objects less than
are actually found outside of $E_{\rm inner}$. However, these 13 clusters
are located so far outside that a region which includes all these objects
cannot be assumed to have an overall constant cluster density. No cluster
pairs are among these outer objects so that they do not need to be included in
our statistics. 55 cluster pairs are found in  $E_{\rm outer}$, but only
$\approx 12 \%$ of them ($\approx 7 \pm 3$ pairs) can be explained due to
chance line-up. Though there is a low probability for a triple system (0.02),
no larger group occurred in our simulations.  

Table~\ref{simgroupstat} also lists the sum of all cluster pairs and groups
that can be expected if the figures for all regions are added up. The last
line of Table~\ref{simgroupstat} gives the group statistics for an entire
artificial LMC, i.e., the experiments for the selected regions are put
together. Again, as was already noticed for Table~\ref{groupstat}, the
statistics show slight differences since some chance pairs and groups are
located across the rims of the selected regions. However, the differences are
much smaller than for the ``real'' LMC (Table~\ref{groupstat}). Comparing the
results with Table~\ref{groupstat} it can be seen that approximately $37 \%$
($288 \pm 15$) of the found 770 cluster pairs can be expected due to spatial
superposition. Approximately $58 \%$ ($211 \pm 12$) of all 366 binary cluster
candidates can be expected statistically. The discrepancy between found and
expected cluster groups increases for larger groups.   

For each region, the number of pairs that can be expected due to random
superposition is much lower than the number of pairs that are actually
found: Between $56 \%$ (in the bar region) and $12 \%$ (in the outer LMC ring)
of all detected pairs can be explained statistically. It seems that the
discrepancy between found and expected pairs also depends on the cluster
density of the region, i.e., in the densest bar region the percentage of pairs
that can be explained statistically is also the highest, while it is the
lowest in the region with the lowest cluster density (the outer ring).  
It is striking that especially large cluster groups with more than four
members scarcely occur in any of the artificial cluster distributions. 

\section{Properties of the multiple clusters}
\label{groupsproperties}

\subsection{Separations between the components of the cluster pairs}
\label{groupssep}

\begin{figure}[]
\centerline{
\includegraphics[width=\hsize]{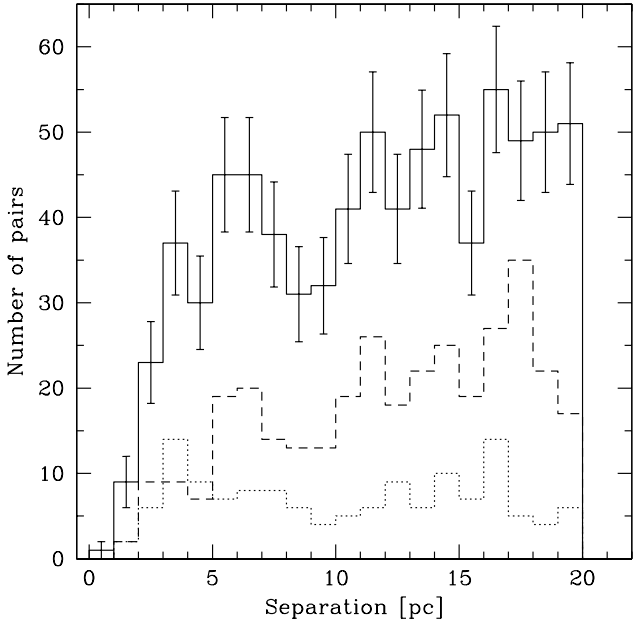}
}
\caption[Distribution of separations for all cluster pairs]{\label{sephisto}
  Distribution of separations for all cluster pairs 
  (solid line) found in the LMC. The distribution seems to be bimodal with
  peaks around approximately 6 and 15 pc and with a decrease around 10 pc. The
  dashed line denotes cluster pairs in which both components have diameters
  larger than 7 pc, while the dotted line represents cluster pairs with
  diameters smaller than 7 pc. The error bars are calculated as
  $\pm\sqrt{N}$}    
\end{figure}

The distribution of the projected centre-to-centre separations of all
LMC cluster
pairs is displayed in Fig.~\ref{sephisto} (solid line). Two peaks around 6 pc
and approximately 15 pc are apparent. The peak around 6 pc as well as the
subsequent decline around 9 pc are independent of binning. The median
separation of the sample is $11.9\pm5.2$ pc, the mode (the most
probable value) is $6.3$ pc. The number of cluster pairs with a
separation of 10 pc and larger increases, but seems to level off or
even decrease again at separations of 18 pc and larger.  

Assuming a uniform distribution of separations we calculate a
median of $43 \pm 9$ clusters per bin. Note that we took into account
only separations between 3 and 20 pc since the number of pairs
observed with very small separations is very low. This might very well
be a selection effect since 
clusters with such small separations might not be detected because they are
overlapping and thus appear as one single object. Figure~\ref{sephisto} shows a
maximum at approx.~6 pc with 45 pairs, a minimum at 9 pc with 31 pairs, and
again a maximum at 17 pc with 55 pairs. The minimum and the second maximum are
significantly below and over the median figure.

To constrain our presumption we performed a KMM (``Kaye's mixture
model'', see Henriksen et al.\ \cite{henriksen}) test. Basically,
mixture modelling is used to detect clusterings in datasets and to assess
their statistical significance. The KMM fits a user-specified number
of Gaussians to a dataset. The algorithm iteratively determines the
best positions of the Gaussians and assigns to each data point a
maximum likelyhood estimate of being a member of the group. It also
compares the fit with the null-assumption, that is a single Gaussian
fit to the dataset, and evaluates the improvement over the
null-assumption using a ``likelyhood ratio test statistic''. The
algorithm is described in detail in Ashman et al.\ (\cite{ashman}).   
The user has to provide as an input the number of data points, an
initial guess for the number of groups, their positions, and sizes. A
great advantage of KMM is that it works on the data themselves and is
not applied to the histogram, thus it is completely independent of
binning and not affected by any subjective visual impression. 

For our first guess, we assumed two distributions with positions
(i.e., the mean of the Gaussians) at 6 and 15 pc, 4 pc as the standard
deviation of the Gaussians, and a mixing proportion of 0.4 and 0.6 for
the two groups. The number of data points
assigned to each group by KMM is 325 and 440 with estimated correct
allocation rates of 0.914 and 0.944 for the two groups. The estimated
overall correct allocation rate is 0.931. The estimated means of the
two groups are 6.644 and 15.442 (close to our assumed positions). The
hypothesis that the distribution can be fitted by a single Gaussian is
rejected with more than 99 \% confidence. 

It might be possible that the underlying distribution is best
described with three Gaussians. Our input guess for this case was
means at 6, 13, and 18 pc, a common standard deviation of 3, and mixing
proportions of 0.4, 0.3, and 0.3. The KMM assigns 239, 239, 287
members to each group, with allocation rates of 0.936, 0.845, and
0.925 and an overall correct allocation rate of 0.903. The KMM
estimated positions are at 5.368, 11.599, and 17.043. Again, the
null-assumption is
rejected with more than 99 \% confidence. Compared to our first, two-Gaussian
guess, the KMM estimate for the overall correct allocation rate is
smaller. We conclude that the distribution is better described with a
two-Gaussian distribution. 

Bhatia \& Hatzidimitriou (\cite{bh}) investigated the separations of their 69
proposed binary clusters and found a bimodal distribution with peaks around 5
and 15 pc, similar to our findings if a two-Gaussian distribution is
assumed. Bhatia et al.\ (\cite{brht}) further
suggested a more uniform distribution for cluster pairs in which both clusters
have diameters larger than 7 pc. However, a uniform distribution for large
clusters can be explained in the following way: the larger the components of a
cluster pair, the larger the probability that both clusters are overlapping
and may not be detected as a cluster pair but as only one single large
cluster. It is likely that the catalogue is not complete concerning cluster
pairs in which both clusters are large but have a small separation.
 
Based on our catalogue of binary and multiple cluster candidates, we
reinvestigated the distribution of separations for cluster pairs in which 
both components have diameters either larger or smaller than 7 pc. The dashed
line in Fig.~\ref{sephisto} denotes pairs consisting of large clusters while
the dotted line stands for pairs with small components. Indeed, the bimodal
distribution is most apparent for small components and seems to be peaked
around approximately 5 and 15 pc, in agreement with the findings of Bhatia et
al.\ (\cite{brht}). For pairs consisting of large clusters, a bimodal
distribution is not as apparent, but cannot be neglected either. We cannot
confirm a uniform distribution of separations for pairs with large
clusters.    

A general increase in the number of pairs as a function of separation
is obvious from 
Fig.~\ref{sephisto}. This increase can be expected because cluster pairs with
larger separations between the components can more easily be detected
than close 
couples of clusters which might overlap and thus ``merge'' into one single
cluster. Besides, the probability of finding another cluster increases
with increasing separation (and thus increasing area).  

On the other hand, for a given separation between cluster pairs, we
expect to find an increase in the number of binary cluster candidates
towards smaller separations since the ``projected'' separations are
smaller than the real one. This might explain the first peak around 6
pc in the distribution of separations. The decrease towards
separations smaller than 6 pc can be expected since clusters with
small separations likely overlap and thus are difficult to detect.
 
Consequently, the dip around 9 -- 10 pc might be interpreted as a
balance between the effects that lead to an increase in the
number of cluster pairs towards either smaller or larger separations.

\subsection{Size distribution}
\label{montecarlosizes}

\begin{figure}[]
\centerline{
\includegraphics[width=\hsize]{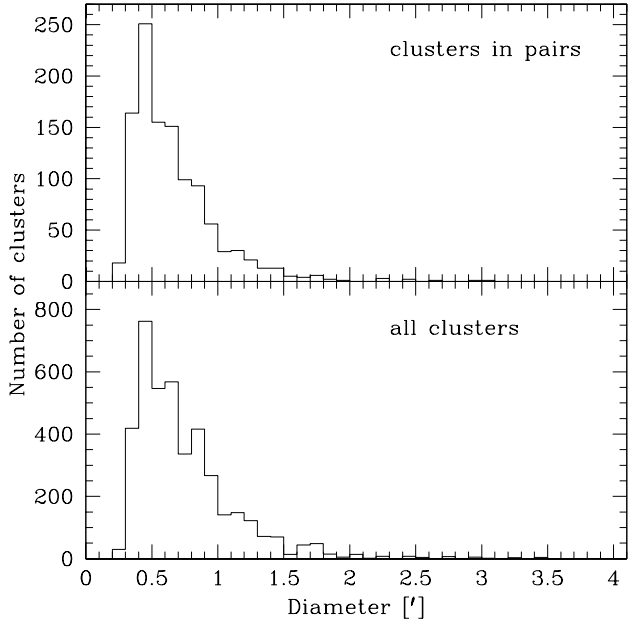}
}
\caption[Size distribution of all clusters and of the clusters involved in
pairs]{\label{diamhisto} Size distribution of the clusters involved in pairs
  (upper diagram) and of all LMC clusters (lower diagram)}    
\end{figure}

The size distribution of clusters that are part of cluster pairs or groups is
displayed in the upper diagram of Fig \ref{diamhisto}. Most components of the
cluster pairs are small. They have diameters between $0.2 \arcmin$ ($\approx
3$ pc) and $1\farcm5$ ($\approx 22$
pc) with a clear peak at $0\farcm45$ ($\approx 6.6$ pc). The median diameter
of the sample is $0\farcm57 \pm 0\farcm26$ ($\approx 8.5 \pm 3.8$ pc), the
mode is at $0\farcm48$ ($\approx 7$ pc). Only a few clusters have diameters
larger than $1\farcm8$ ($\approx 26$ pc).   
However, in spite of our selection criterion of a separation of 20 pc, we
still find three clusters with diameters larger than 40 pc ($2\farcm7$).  
This means that their companion cluster is embedded within their
circumference. These clusters are NGC\,1850 (or BRHT\,5\,a) with its
companions NGC\,1850\,A and BRHT\,5\,b (or H88-159), and NGC\,2214 which
appears in the BSDO catalogue as two entries, namely
NGC\,2214\,w and NGC\,2214\,e.  

The lower diagram in Fig.~\ref{diamhisto} shows the diameter distribution for
all clusters found in the BSDO catalogue. Again, most
clusters are rather small with a peak at $0\farcm45$ or $\approx 6.6$ pc.
The median diameter of the entire cluster sample is $0\farcm62 \pm 0\farcm41$
($\approx 9 \pm 6$ pc) and the mode is $0\farcm55$ or $\approx 8$ pc.

Both distributions (upper and lower figure) are qualitatively very similar. 

\begin{figure}[]
\centerline{
\includegraphics[width=\hsize]{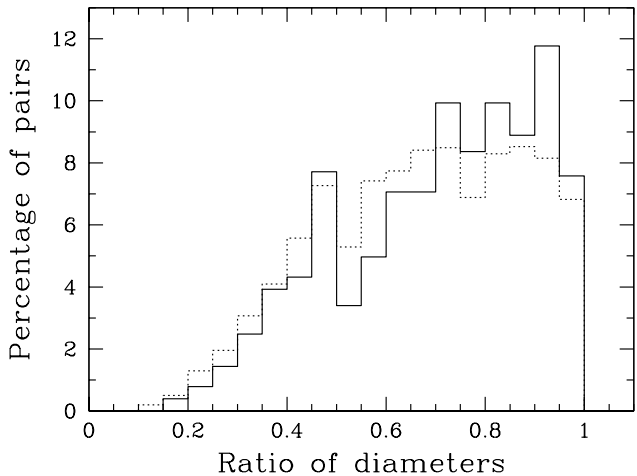}
}
\caption[Ratio of the diameters of the components of the cluster
pairs]{\label{ratio} Diameter ratio of the components forming a cluster
  pair (solid line). The number of pairs increases with increasing ratio. The
  dotted line represents the distribution of ratios for scrambled
  diameters. See Sect.~\ref{montecarlosizes} for the details}  
\end{figure}

The normalized ratio of the diameters of clusters that form a pair are plotted
in Fig.~\ref{ratio}. The median ratio of the sample is $0.73 \pm 0.2$.
The number of cluster pairs increases towards a size ratio of 0.5,
but drops at a ratio larger than 0.5 and lower or equal than 0.55, and then
increases again towards a ratio of 1. The number of pairs increases with
larger ratios, which might indicate that binary clusters tend to form with
components of similar sizes.   

The dotted line in Fig.~\ref{ratio} represents the size ratio of cluster pairs
if all diameters are mixed and then randomly assigned to the pair members. To
get reliable statistics we repeated this procedure 100 times. The number of
pairs increases with increasing ratio, but seems to decrease again at ratios
larger than 0.75, which confirms the impression that ``true'' binary clusters
tend to form with components of similar sizes. Again, there is a peak at 0.5
and a following dip at ratios slightly larger than 0.5, though not as
prominent as in the distribution of found ratios (solid line). However, a
uniform distribution is not expected for statistical reasons: the diameters
of the clusters in the BSDO catalogue are given in arc
minutes in steps of $0\farcm05$, i.e., the smallest diameter is $0\farcm25$,
the next one $0\farcm3$ and so on. Since we consider mean diameters, we
obtain discrete values with an increment of $0\farcm025$. 
This means that some ratios are more probable than other ones, namely the unit
fractions, which includes a ratio of $0.5 = 1/2$, while other ratios might
result only few times in the distribution. For example, a ratio of
$34/35$ can only 
result from three combinations of diameters in the given distribution of
diameters, namely if both components of the pair have diameters of $0\farcm85$
and $0\farcm875$, or $1\farcm7$ and $1\farcm75$, or $2\farcm55$ and
$2\farcm625$. In the real ratio distribution it occurs only once for
$0.85/0.875$. This explains the peak at 0.5 as one of the very likely
ratios in the distribution.  
   
However, in general the distribution of the found ratios and the distribution
of the ratios for scrambled diameters agree well with each other, though there
might be a tendency of the real binary cluster candidates to form more pairs
with components of similar sizes.

\subsection{Spatial distribution of the cluster pairs and groups}
\label{sectpairslocation}

\begin{figure}[]
\centerline{
\includegraphics[width=\hsize]{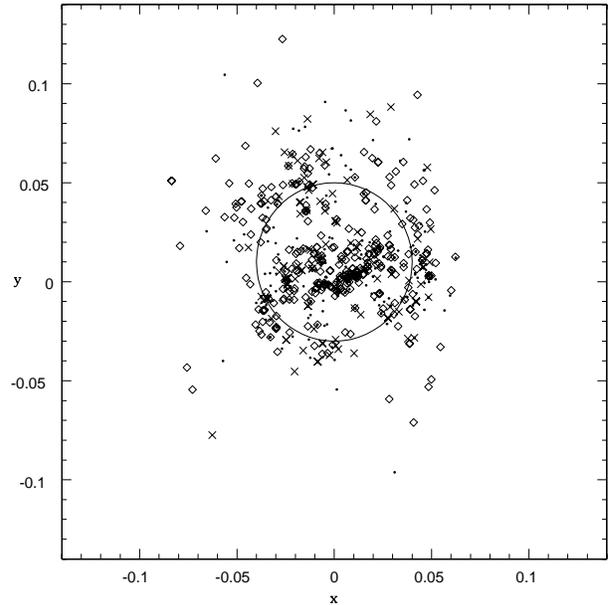}
}
\caption[Location of the cluster pairs]{\label{pairslocation} Location of all
  cluster pairs in the LMC. Regions of increased pair density correlate with
  the distribution of star clusters in general. Diamonds denote cluster pairs
  with both components larger than 7 pc, crosses stand for pairs in which the
  clusters have smaller diameters. Dots represent cluster couples in which one
  component is larger and the other one is smaller than 7 pc. The circle
  marks the boundary between the inner and outer LMC for which we compared
  the ratio of pairs with only large or only small components. See
  Sect.~\ref{sectpairslocation} for the details}   
\end{figure}

Figure~\ref{pairslocation} represents the location of all cluster pairs found
in the LMC. The distribution of all pairs reflects the dense bar region and
the region around the bar ($E_{\rm bar}$). The pair density drops
considerably in the outer LMC region. Altogether, the distribution of cluster
pairs is very similar to the distribution of clusters in general and there are
no regions of increased pair density that do not correlate with the
distribution of clusters.  

Bhatia et al.\ (\cite{brht}) suggested that pairs with small clusters are
predominantly found outside the central region of the LMC. However, they
caution that this effect might also be due to the increasing incompleteness
for small clusters in their data towards the crowded inner LMC. We
reinvestigated the distribution for cluster pairs in which both components are
either larger (diamonds in Fig.~\ref{pairslocation}) or smaller (crosses in
Fig.~\ref{pairslocation}) than 7 pc. Most cluster pairs have 
large components, in total 336 pairs. 136 pairs have only small clusters,
and the remaining 293 couples have a small as well as a large component. It
seems that in the outer LMC comparably more pairs with large clusters can be
found than pairs with small components. The ratio of pairs with only large
components and pairs with only small ones is $336/136=2.46$. If only pairs in
the inner parts of the LMC (marked with a circle in Fig.~\ref{pairslocation})
are considered, the ratio is $200/75=2.67$, for the outer region it is
$136/61=2.23$. This means that in the outer as well as in the inner
LMC, more pairs with only large components than pairs with only small
clusters can be found, however, in the outer LMC we find
proportionally more pairs with only small clusters compared to the
inner LMC. However, in total numbers most of the pairs with only small
components are found in the inner parts of the LMC, opposite to the
suggestion of Bhatia et al.\ (\cite{brht}). 

In general, the distributions seem to follow the distribution of cluster
pairs and we do not see regions that are primarily populated with pairs of a
specific ``type'' that differ from the general distribution of clusters. We
cannot confirm the accumulation of pairs with only small clusters in the outer
LMC region as suggested by Bhatia et al.\ (\cite{brht}). Their finding is
likely an effect of the incompleteness of their data (they considered
69 binary cluster candidates whereas our sample includes 765 cluster pairs).

\subsection{Ages of the binary and multiple cluster candidates}
\label{pairsages}

We have searched for ages of the binary and multiple cluster candidates in the
literature. Age information is available only for a fraction of all the
clusters in our catalogue. It turned out that out of a total of 473 groups
only 186 groups have age information available, and the information is
complete for all group components only for a fraction of these
groups. In total,
we found ages for only 306 clusters, which are $\approx 27 \%$ of 
the 1126 clusters that form pairs and groups. The most fruitful sources were
the publications of Bica et al.\ (\cite{bcdsp}), who estimated ages from
integrated $UV$ photometry, and of the OGLE group, 
who fitted isochrones to CMDs (Pietrzy\'{n}ski \&
Udalski \cite{pu_lmc}). 
All results are summarized in Table~\ref{groupscatalogue} where we
also give the 
corresponding references. This catalogue contains all binary and multiple
cluster candidates found in the entire LMC, based on the BSDO
catalogue (see Sect.~\ref{groups_montecarlo} where we noted 
the different number of groups found in the entire LMC and the sum of the
groups found in all regions separately).

\begin{figure}[t]
\centerline{
\includegraphics[width=\hsize]{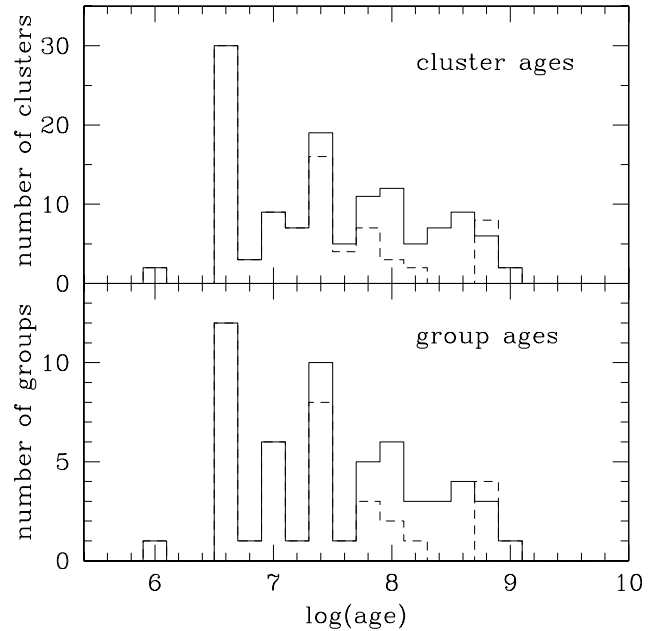}
}
\caption[Ages of the cluster groups]{\label{agehisto} Upper diagram: age
  distribution of all clusters found in groups and for which age information
  is available. Lower diagram: age distribution for the groups for which the
  members have ages similar enough to agree with a common origin. The ages of
  the group members are averaged and the mean age is assigned to the group and
  plotted in this figure. The dashed line in both diagrams denotes the age
  distribution if ages derived by Pietrzy\'{n}ski \& Udalski (\cite{pu_lmc})
  are not considered. As can be seen, the OGLE ages (Pietrzy\'{n}ski
  \& Udalski \cite{pu_lmc}) are the major contribution to old clusters
  and groups}    
\end{figure}

In Fig.~\ref{agehisto} we plotted a histogram of the age distribution for our
binary and multiple cluster candidates. If more than one age was determined
for a cluster we averaged the values. However, if the ages 
found by various authors differ considerably we adopt the value found in the
most recent studies since the methods of age determinations have improved
in the recent years, e.g., ages derived from isochrone fitting to CMDs
based on CCD photometry are generally considered the most reliable and
accurate age determinations.  

An example is NGC\,1775 for which Bica  et al.\ (\cite{bcdsp})
estimated an age 
of 70 -- 200 Myr while Kontizas et al.\ (\cite{kkm}) stated that the stars in
NGC\,1775 are too faint for their detection limit and thus suggested an age
larger than 600 Myr. Since Bica  et al.\ (\cite{bcdsp}) did not report
detection problems for this object, we adopt a mean age of 135 Myr for this
cluster to be plotted in Fig.~\ref{agehisto}. 

\begin{figure}[]
\centerline{
\includegraphics[width=\hsize]{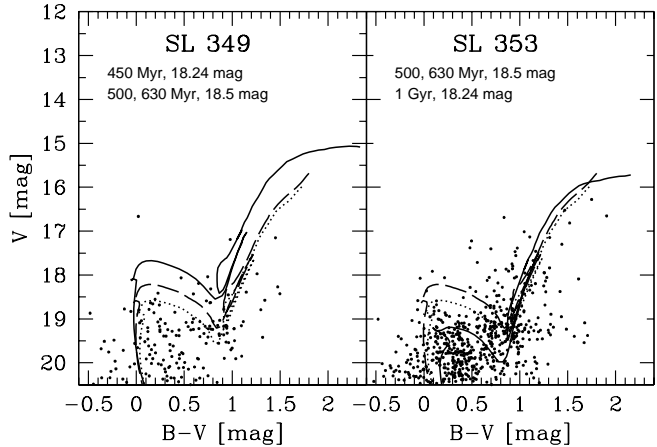}
}
\caption[SL\,353 \& SL\,349: Comparison of the OGLE isochrone fit and
ours]{\label{ogle_sl353} Comparison of the OGLE isochrone fit and ours. The
  data are from the OGLE Internet archive
  ({\tt ftp://bulge.princeton.edu/ogle/ogle2/clusters/lmc/}). Overplotted are
  the Padua isochrones suggested by Pietrzy\'{n}ski \& Udalski (\cite{pu_lmc})
  that are based on a distance modulus of 18.24 mag and lead to an age
  of 1 Gyr for SL\,353 and 450 Myr for SL\,349 (solid lines). It seems
  that SL\,349 is older and SL\,353 younger than these suggested
  ages. Geneva isochrones that are based on a distance modulus of 18.5
  mag and that represent an age of 500 (dashed line) and 630 Myr
  (dotted line) are also plotted and give a better fit. From our
  isochrone fitting we derived an age of $550 \pm 100$ Myr (see
  Dieball et al.\ \cite{dgt})}       
\end{figure}

An example for which ages derived from isochrone fitting is available
is SL\,353 \& SL\,349: CCD based CMDs were investigated by Dieball et
al.\ (\cite{dgt}) and by Vallenari et al.\ (\cite{vbc}) and both
studies agree with ages of 550 Myr for both clusters. Bica et al.\
(\cite{bcdsp}) derived an age of 1.4 Gyr from integrated
colours. However, Geisler et al.\ (\cite{geisler97}) pointed out that a few 
bright stars can influence the age determination based on integrated
photometry, making the result dependent on the chosen aperture. 
Pietrzy\'{n}ski \& Udalski (\cite{pu_lmc}) fitted isochrones to CMDs
and suggested an age of 1 Gyr for SL\,353 and 450 Myr for
SL\,349. These authors used a distance modulus of 18.24 mag and fitted
isochrones based on the stellar models of the Padua group (Bertelli et
al.\ \cite{bertelli}), whereas we use a modulus of 18.5 mag and
isochrones based 
on the Geneva models (Schaerer et al.\ \cite{smms}). However, Vallenari et
al.\ (\cite{vbc}) also used the Padua isochrones and their results agree with
ours. The smaller distance modulus of 18.24 mag would lead to larger ages,
this cannot explain the smaller age that Pietrzy\'{n}ski \& Udalski
(\cite{pu_lmc}) found for SL\,349 and the age difference suggested for 
the cluster pair. In Fig.~\ref{ogle_sl353} we compare their isochrone fit with
ours. It seems that their suggested age for SL\,349 is too young while the age
for SL\,353 seems to be too old to give a good fit. Isochrones representing
the ages we adopted for this cluster pair are also plotted (see
Dieball et al.\ \cite{dgt} for the details).

In cases where several consistent age determinations are available,
but one value differs from the others, we omit the ``outlier'' and
average the other results. This is the case, e.g., for SL\,229. For
this cluster Fujimoto \& Kumai (\cite{fk}) derived an age of 460 Myr,
Bica et al.\ (\cite{bcdsp}) suggested an age of 200 -- 400 Myr, and
Pietrzy\'{n}ski \& Udalski (\cite{pu_lmc}) 220 Myr, however, Kontizas
et al.\ (\cite{kkm}) proposed 6 -- 80 Myr. We adopt a mean of 330
Myr. For the companion cluster SL\,230 the age 
determinations agree better: 74 Myr (Fujimoto \& Kumai \cite{fk}), 20 Myr
(Bica et al.\ \cite{bcdsp}), 43 Myr (Kontizas et al.\ \cite{kkm}), and 140 Myr
(Pietrzy\'{n}ski \& Udalski \cite{pu_lmc}). We adopt 70 Myr, which agrees
with Fujimoto \& Kumai (\cite{fk}) and Kontizas et al.\ (\cite{kkm}), but is a
higher value than suggested by Bica et al.\ (\cite{bcdsp}) and lower than
suggested by Pietrzy\'{n}ski \& Udalski (\cite{pu_lmc}). 

In this way different ages for the same cluster are averaged to a mean
age, however, the main information, which is if the clusters of a group
have ages similar enough to agree with a common formation or not, is
obtained in all cases. 

In any case, in Table~\ref{groupscatalogue} we list all results found for each
object. 

In some cases no ages could be found for the specific clusters of a group, but
an age determination of the surroundings, e.g., the association the
clusters are embedded in, is available and is adopted for the plot in
Fig.~\ref{agehisto}. For example, we assume an age of 5 Myr for BSDL\,1437 \&
HD\,269443, which are both embedded in LMC\,N\,44\,D for which Bica et al.\
(\cite{bcdsp}) derived a mean age of 5 Myr. In such a case a congruous remark
is made in Table~\ref{groupscatalogue}.  

For only 96 groups age information is available for more than one cluster,
which allows a closer look at the age structure of the specific group, though
ages are rarely found for ``all'' clusters of a group. 

If clusters have formed from the same GMC, they should be coeval or have
age differences that are small enough to agree with a common
formation, i.e., the age difference must be smaller than the maximum
life time of a GMC. Fujimoto \& Kumai (\cite{fk}) suggested that the
life time of a protocluster gas cloud is of the
order of a few 10 Myr. However, more recently Fukui et al.\
(\cite{fukui}) and Yamaguchi et al.\ (\cite{yama}) suggested that the
life time of a GMC is of the order of only a few Myrs:

Fukui et al.\ (\cite{fukui}) conducted a CO survey of the LMC,
catalogued the CO clouds, and correlated their positions with all
clusters listed in the Bica et al.\ (\cite{bcdsp}) catalogue, which
contains also age estimates for the clusters. Fukui et al.\ (\cite{fukui})
found a significant correlation of the positions of the youngest
clusters (SWB\,0, age $\leq 10$ Myr) with nearby CO clouds. In contrast, the
location of older clusters (SWB\,II -- SWB\,VII) with respect to
nearby CO clouds was found to be consistent with a random
distribution, i.e., they can easily be explained as line-of-sight
chance superpositions. The authors suggested that star clusters are
formed rapidly in a few Myr after cloud formation and that the cloud
dissipates  quickly on a time scale of 6 Myr.
More recently, Yamaguchi et al.\ (\cite{yama}) suggested that the GMCs
actively form star clusters for about 4 Myr, and that they are completely
dissipated due to the winds and supernova explosions of massive stars
within the following 6 Myr (Yamaguchi et al.\ \cite{yama}, their Table~5).  
Fukui et al.\ (\cite{fukui}) found that approximately 30~\% of the
young clusters with ages $< 10$ Myr are located within 130 pc from the
surviving CO clouds.  
  
This implies that the time scale for the joint formation of a cluster
pair that fulfill our criterion of 20 pc must be on average less than
10 Myr. This results in a rather stringent age criterion for true
binary clusters.

On the other hand, we need to take into account that for clusters of
an age of $\approx 100$ Myr and older the age resolution is worse than
10 Myr and continues to decrease. Hence it seems to be justified to
consider two components of a potential binary cluster coeval when
their ages agree within the uncertainties of their age determination.

In 57 groups at least two clusters appear to be either coeval or have
ages similar enough to agree with a common formation in the same GMC,
i.e., the age differences are smaller than 10 Myr. 
To be able to plot the group ages (see Fig.~\ref{agehisto}, lower
diagram) we have averaged the ages of the group members and assigned a mean
age to the corresponding group. For some of the older
clusters, the age difference inside the group can be larger than 10
Myr, but still within the errors the group components agree with the
same age (see text above). This is the case, e.g., for group no.~206 where
Pietrzy\'{n}ski \& Udalski (\cite{pu_lmc}) derived an age of 500 Myr
for KMK\,88-49. For NGC\,1938, Pietrzy\'{n}ski \& Udalski
(\cite{pu_lmc}) found an age of 355 Myr, Fujimoto \& Kumai
(\cite{fk}) estimated an age of 550 Myr, Bica et al.\ (\cite{bcdsp})
suggested 200 -- 400 Myr, Kontizas et al.\ (\cite{kkm}) suggested an
age $>$600 Myr. We adopt a mean of $450\pm140$ Myr for
NGC\,1938. Within the errors, both clusters, NGC\,1938 and KMK\,88-49,
agree well with a common formation from the same GMC. For the third
component of this group, NGC\,1939, all age estimates lead to higher
ages of 7 Gyr (Fujimoto \& Kumai \cite{fk}), 5 -- 16 Gyr (Bica et
al.\ \cite{bcdsp}), $>$600 Myr (Kontizas et al.\ \cite{kkm}), and 1 Gyr
(Pietrzy\'{n}ski \& Udalski \cite{pu_lmc}). We adopt a mean of 5 Gyr.
It is clear that NGC\,1939 is considerably older than the other two
clusters of this group and cannot have formed together with the other
two components. 
In general, the error of the age determination is the larger the older
the cluster is.
The groups that for this reason show somewhat higher internal age
differences than our selection criterion of 10 Myr are nos.~90, 94, 124,
135, 180, 184, 206, 211, 243, 428, and 456. 
In the remaining 39 groups the age difference(s) found well exceed 10
Myr (also when the errors in the age determination are considered)
which is more than the maximum life time of protocluster gas clouds
(Fukui et al.\ \cite{fukui}, Yamaguchi et al.\ \cite{yama}). As a
result, these clusters cannot have a common origin.  

\begin{figure}[]
\centerline{
\epsfig{file=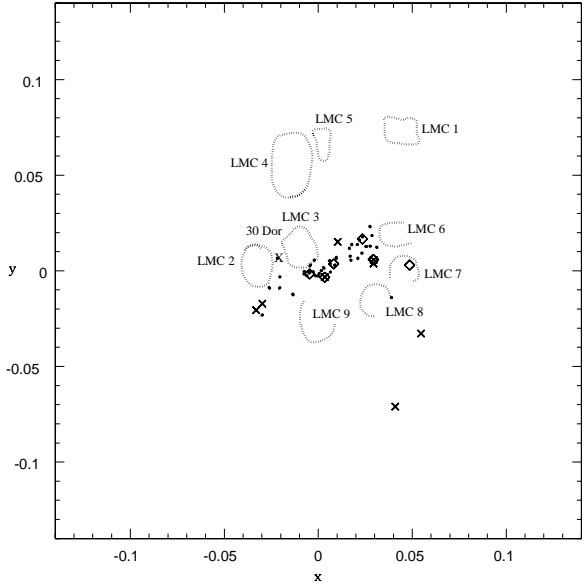, width=8.5cm, clip=}
}
\caption[Location of old cluster groups]{\label{groupoldloc} 
  Location of cluster groups that are older than 300 Myr (crosses) or have
  large internal age difference which do not agree with a common formation of
  the group components (dots). Diamonds denote cluster groups that
  comprise a binary or triple cluster and one or two additional
  clusters whose ages indicate that they did not form with the binary
  or triple system. The location of the supergiant shells and
  30\,Doradus are sketched as an orientation guide. As can be seen, the
  majority of the groups are located in the bar region}    
\end{figure}

In Fig.~\ref{agehisto}, upper diagram, the ages of all clusters with available
age information are plotted. As can be seen, the clusters are predominantly
young (a few 10 Myr to 100 Myr) or very young (a few Myr) with significant
peaks at 4 Myr, 25 Myr, and 100 Myr. Smaller peaks are at 10 Myr and
400 Myr. Only a few clusters are older than
1 Gyr, and if so, their companion cluster(s) is of a different (younger) age
which makes it likely that the specific group appears close on the sky only
due to projection effects. An exception is group no.~11 where both clusters
have an age of 1.2 Gyr. 
We agree with Pietrzy\'{n}ski \& Udalski (\cite{pu}) that most clusters are
younger than 300 Myr. The peak at 100 Myr might be explained by a close
encounter of both Magellanic Clouds roughly 200 Myr ago that triggered
star and cluster formation (Gardiner et
al.\ \cite{gardiner}). However, our age distribution for the group components
differs from the one presented by Pietrzy\'{n}ski \& Udalski (\cite{pu}). The
pronounced peaks at 4 and 25 Myr are missing in Pietrzy\'{n}ski \& Udalski's
(\cite{pu}) age distribution,
which is due to the fact that these authors investigated only the
central part of the LMC whereas we study the whole LMC area. Most of
the clusters younger than 30 Myr are located outside the LMC bar,
whereas older clusters are concentrated towards the bar region (see
e.g.\ Fig.\ \ref{groupoldloc}). 
In addition, the smaller distance modulus of 18.24 mag used by
Pietrzy\'{n}ski \& Udalski (\cite{pu_lmc}) also leads to higher ages. The
dashed line in Fig.~\ref{agehisto} shows the age distribution of the clusters
and groups if the OGLE ages (Pietrzy\'{n}ski \& Udalski \cite{pu_lmc})
are not considered. As can be seen, the OGLE ages are the major
contribution to clusters with ages of 100 Myr or older.   
 
The lower diagram in Fig.~\ref{agehisto} shows the age distribution of
the cluster groups for which 
the members are coeval or have ages similar enough to agree with a common
origin. Again, most groups are found to be quite young and only 8 groups
(nos.~4, 11, 83, 84, 174, 206, 408, 428 in Table~\ref{groupscatalogue})
are older than 300 Myr. However, there might also be selection effects in the
sense that older cluster groups might not be detected because the clusters are
too faint, or old systems do not exist anymore because they are already
dissipated or merged into a single cluster. The dashed line denotes the group
age distribution if the OGLE ages are not considered. In this case only 5
groups (nos.~4, 11, 174, 408, 428) are older than 300 Myr. Again, the OGLE
ages contribute primarily to the groups with ages of 100 Myr or
older. The inclusion of the OGLE ages also changes the mean group ages
for some of the groups (namely group nos.~408, 428), which
explains the smaller count at 630 Myr ($\rm{log}~t = 8.8$)
compared to the group mean age distribution if the OGLE ages are not
considered.

In Fig.~\ref{groupoldloc} the location of the old groups (older than
300 Myr, plotted as crosses) and groups with internal age differences
which do not agree with a common formation of  the group components 
(indicated as dots) are plotted. As can be seen,
most of these groups are located in the dense bar region and thus can
easily be explained with chance superpositions. 

Group no.~83 comprises a binary cluster candidate, BRHT\,3b and
KMK\,88-4. For both clusters, Pietrzy\'{n}ski \& Udalski
(\cite{pu_lmc}) derived an age of 630 Myr. Two more clusters can be
found in this cluster group: H\,88-107 (710 Myr, Pietrzy\'{n}ski \&
Udalski \cite{pu_lmc}) and NGC\,1830. For the latter cluster we adopt
a mean age of 275 Myr. H\,88-107 is too old to agree with a common
formation together with the binary cluster candidate, and NGC\,1830 is
too young. These two clusters are most probably chance
superpositions. Also groups nos.~ 14, 110, 180, 206, and 258 contain a
binary or triple cluster candidate and one ore more components that do
not agree with a common formation. These groups are plotted as
diamonds in Fig.~\ref{groupoldloc}. Nearly all of them are located in the
dense bar region.

\begin{figure}[]
\centerline{
\includegraphics[width=\hsize]{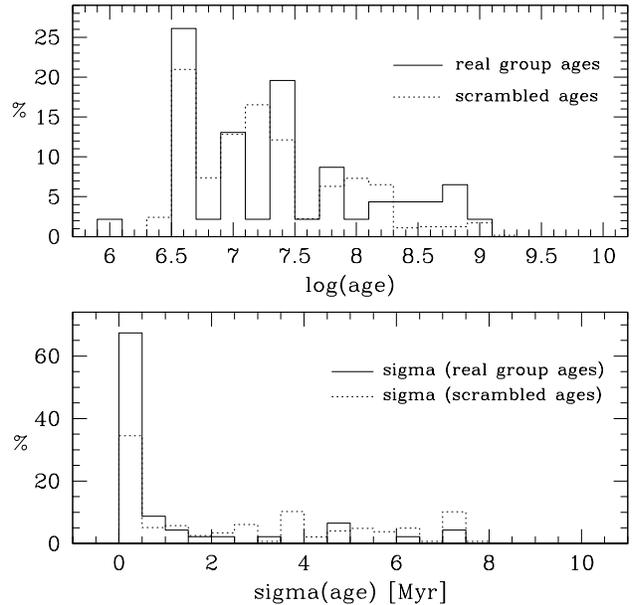}
}
\caption[Group ages resulting from scrambled member ages and their internal
age scatter]{\label{simgroupage}
  Upper diagram: distribution of group ages if all the cluster ages are
  scrambled and randomly assigned to the group members (dotted line). The solid
  line represents the real group age distribution (see Fig.~\ref{agehisto}),
  but without groups nos.~90, 94, 124, 135, 180, 184, 206, 211, 243,
  428, and 456. Lower diagram: distribution of internal age deviations
  for the real group ages (solid line) and for the group ages based on
  scrambled cluster ages (dotted  line). The real group age
  distribution shows pronounced peaks and has smaller internal
  scatter than the distribution based on randomly mixed member ages}     
\end{figure}

The upper diagram of Fig.~\ref{simgroupage} shows the distribution of group
ages if all ages for the clusters are scrambled and then randomly assigned to
the group members for which the age information was available. In this way,
the groups' mean ages change, but also the  
number of groups which a mean age can be ascribed to varies. We repeated this
procedure 100 times to get reliable statistics. On average, $12.9\pm2.7$
groups per run have two or more clusters that are either coeval or
have age differences small enough to agree with a common formation so that a
mean age could be assigned to the corresponding group. The remaining groups
with more than one member age have internal age differences larger than 10
Myr. In a few cases, a group with four or more members could be subdivided into
two groups with two (or more) coeval or nearly coeval clusters. Each subgroup
was counted as a single group. 
In the real age distribution 46 groups were found that have internal age
differences $\le 10$ Myr. Please note that this number does not include the
groups nos.~90, 94, 124, 135, 180, 184, 206, 211, 243, 428, and 456. These
groups show larger internal age differences than our selection
criterion of $\Delta t < 10$ Myr but are included in
Fig.~\ref{agehisto}. 
In a way, these groups are borderline cases to our selection
criterion, see the text above. For Fig.~\ref{simgroupage}, we use the
stringent selection criterion that can easily be applied to the groups
with scrambled cluster ages and thus makes a direct comparison possible.
However, the number of real groups that match this strict selection
criterion is significantly larger (more than 3.5 times) than the
expected number of groups if the clusters' ages are randomly
distributed (46 groups found compared to 12.9 groups expected). 
The peaks at 100 Myr and 400 Myr in Fig.~\ref{agehisto} are not seen
in Fig.~\ref{simgroupage} for the real group age distribution. If the
``borderline cases'' (see above) are not considered, then the age
distribution shows peaks at 4 Myr, 10 Myr, 25 Myr, 63 Myr and 630
Myr, i.e., the two peaks at the older ages are shifted to the next younger
and older bin, respectively.  
The distribution of group ages resulting from scrambled member ages in
Fig.~\ref{simgroupage} is normalized so that the ordinate gives the number of
groups in percent. For comparison the distribution of the real group ages is
also plotted (solid line). As can be seen, the distribution based on the
scrambled cluster ages (dotted line) is smoother with peaks at 4 Myr,
16 Myr, and 100 Myr. Only few groups are older than 200 Myr.
 
The lower diagram in Fig.~\ref{simgroupage} shows the deviation in age for the
group ages. The solid line represents the deviation for the real group
ages: $\approx 70 \%$ of all groups show no or a very small age deviation
(smaller than 0.5 Myr), and the mean sigma is about $1.0\pm2.0$ Myr.  
In contrast, the deviation for the group ages based on the scrambled
cluster ages (dotted line) is smoother, i.e., fewer groups have
smaller ($\approx 35 \%$) and
more groups have larger deviations than 0.5 Myr when compared to the real group
age deviations. The mean deviation for the group ages based on the mixed
cluster ages is $2.7\pm2.6$ Myr and thus larger than the mean
internal scatter for the real group ages.  
If only groups are considered with $\sigma(t)>0.5$ Myr, the mean sigma is
$3.2\pm2.4$ Myr for the real distribution and $4.1\pm2.2$ Myr for the
groups with scrambled cluster ages.

\begin{figure}[]
\centerline{
\includegraphics[width=\hsize]{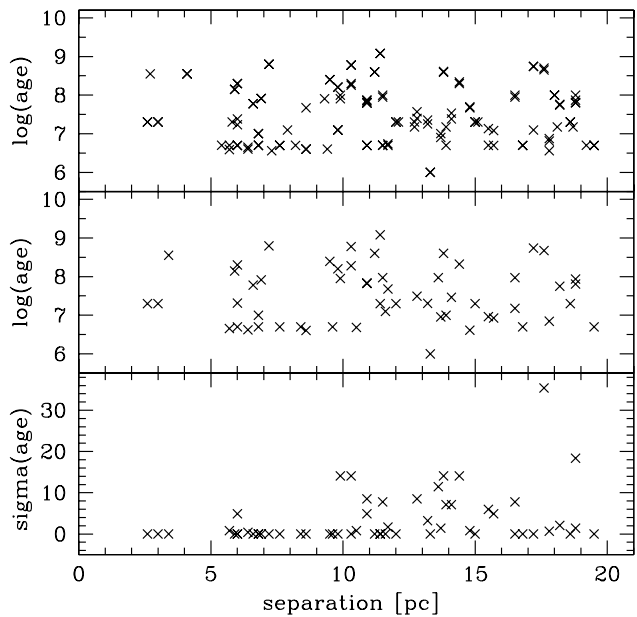}
}
\caption[Ages of the multiple cluster components versus
separation]{\label{agesep} Upper diagram: ages of the multiple cluster
  components versus their separations. Middle diagram: group ages versus their
  internal mean separation. Lower diagram: internal age scatter (in Myr) for
  the groups versus their internal mean separation}    
\end{figure}

In Fig.~\ref{agesep} we plotted the ages found for the clusters versus the
separations that the clusters have within the groups (upper diagram),
and the group ages versus their internal mean separations (middle
diagram). 
No correlation can be seen from Fig.~\ref{agesep}. Thus we cannot draw
any conclusions whether older 
groups or group components tend to have larger (mean) separations,
which would indicate that the components of the multiple cluster are
drifting apart, or whether older clusters have smaller separations, 
which might indicate that the system will undergo a merging process. Both
processes could be equally likely, which might explain why we see no tendency
towards larger or smaller separations. 
 
The lower diagram of Fig.~\ref{agesep} displays the groups' internal age
scatter versus their internal mean separation. There might be a
tendency towards 
larger age scatter with larger mean separations (which indicates larger
groups), but if so, it is only weak. 
Note that we took ``all'' groups into account whose components agree
within the errors of their age determination with a common formation,
i.e., groups nos.~90, 94, 124, 135, 180, 184, 206, 211, 243, 428, and
456 are included. The data point at $\sigma({\rm age}) \approx
36$~Myr belongs to group no.~206 whose components differ in age by 50 Myr,
but considering their age of 500 Myr and 450 Myr, both clusters agree
within the errors with a common formation. 
Efremov \& Elmegreen (\cite{ee}) proposed
that close clusters in pairs have more similar ages. The pairs' average age
differences increase with increasing separations between the
clusters. However, the authors do not restrict their study to binary cluster
candidates that have separations of 20 pc ($1\farcm4$) or smaller. Indeed,
their Fig.~1 seems to indicate that only very few binary cluster candidates
but pairs with much larger distances were considered. However, we cannot
confirm the strong tendency suggested by Efremov \& Elmegreen (\cite{ee}).   

\section{Summary and conclusions}
\label{summary_montecarlo}

We investigated the BSDO catalogue and provide a new
catalogue of all binary and multiple cluster candidates found in the LMC. The
catalogue is presented in Table~\ref{groupscatalogue}. Age information
available in the literature is also given. We found in total 473 multiple
cluster candidates. The separations between the clusters' centres are $\le
1\farcm4$ corresponding to 20 pc (assuming a distance modulus of 18.5 mag).

We performed a statistical study of cluster pairs and groups. For this purpose
we distinguished between regions of different cluster densities in the LMC. 

Vallenari et al.\ (\cite{vbc}) and Leon et al.\ (\cite{lbv}) proposed that the
encounter rate in large cluster groups is higher so that binary clusters can
be formed through tidal capture. Such a scenario might explain large age
differences between cluster pair components. For each selected region
we calculated the 
encounter rate of star clusters. However, we found that the probabilities for
cluster encounters are universally very low. In addition, the probability of
{\it tidal capture} depends on further constraints which will not be fulfilled
during every encounter. Thus we conclude that it seems unlikely that a
significant number of young pairs may have formed in such a scenario. 

We counted the number of all cluster pairs and groups found in the selected
areas. In order to check how many of these multiple cluster
candidates can be expected statistically due to chance line-up, we performed
Monte Carlo experiments for each region to produce artificial cluster
distributions which are compared with the real LMC cluster
distribution. For all selected regions, the number of chance pairs in our
simulations is much lower than the quantity of cluster pairs found: Between
$56 \%$ (in the bar region) and $12 \%$ (in the outer LMC ring) of all
detected pairs can be explained statistically. Especially large cluster groups
with more than four members hardly occur in the artificial cluster
distributions. A significant number of the cluster pairs and groups cannot be
explained with chance superposition and thus might represent ``true''
binary and multiple clusters in the sense of common origin and/or
physical interaction. 

We studied the properties of the multiple cluster candidates:

In the distribution of the centre-to-centre separations of the cluster pairs
two peaks around 6 pc and 15 pc are apparent. This bimodal
distribution is more apparent for cluster pairs in which both components have
diameters smaller than 7 pc, but cannot be neglected for pairs consisting of
larger clusters. We cannot confirm a uniform distribution of separations for
pairs with large clusters as suggested by Bhatia et al.\ (\cite{brht}). Around
separations of 9 -- 10 pc, the number of cluster pairs is depleted.
This dip might be interpreted as a balance between the effects that lead
to an increase in the number of cluster pairs towards either smaller
(due to projection effects) or larger separations (pairs with larger
separations are more easily detected).  
  
The size distribution of the group components shows a peak at $0\farcm45$
($\approx 6.6$ pc). Most clusters involved in pairs or groups are small and
only few clusters have diameters larger than $1\farcm8$ (26 pc). The size
distribution for group components is very similar to the size distribution for
all LMC clusters. It seems that binary clusters tend to form with
components of similar size. 

The spatial distribution of the multiple cluster candidates coincides with the
distribution of clusters in general. 

Only for a fraction ($\approx 27 \%$) of the clusters that form binary and
multiple cluster candidates age information is available, and for only 96
groups ages are known for more than one cluster so that the age structure of
the specific group can be examined. For 57 groups the members appear to be
either coeval or have ages similar enough to agree, within the errors
of the age determination, with a common formation in the same GMC,
i.e., the age differences are 10 Myr at maximum (Fukui et al.\
\cite{fukui}, Yamaguchi et al.\ \cite{yama}). 
The remaining 39 groups have internal age differences which make a
common origin of the components unlikely. 

The clusters involved in pairs or groups are found to be predominantly
young. The age distribution shows peaks at 4 Myr, 25 Myr and 100 Myr. Our
findings differ from Pietrzy\'{n}ski \& Udalski (\cite{pu}) in a way that the
two peaks at the younger ages are missing in their age distribution. This
is due to the fact that these authors investigated only a part of the
LMC and used also a smaller distance modulus that leads to higher ages in
general.   

We scrambled the ages of the groups components and then randomly assigned them
to the group members. On average, $12.9\pm2.7$ groups with internal
age differences $\le 10$ Myr can be expected, however, 46
groups with internal age differences $\le 10$ Myr can be found
in the real distribution (note that the borderline cases are not
considered in this number, see Sect.~\ref{pairsages}), a number
significantly larger than the expected one. Also, the group age
distribution for scrambled member ages is smoother than the real one,
and the internal age scatter is significantly larger for the groups
with random member ages.  

No correlation was found between the groups' ages and their internal mean 
separation. However, there might be a weak tendency towards larger internal
age scatter with larger internal separations (indicating larger groups) but a
strong tendency as suggested by Efremov \& Elmegreen (\cite{ee}) cannot be
confirmed. 

Most multiple cluster candidates are found to be younger than 300 Myr. A
larger number of old cluster groups or of groups with different ages for the
components are not found. A formation scenario through tidal capture
is not only unlikely due to the 
very low probability of tidal capture (even in the dense bar region), but the
few old groups and the groups with large internal age differences can easily
be explained with projection effects, especially since the majority of these
groups are located in the dense bar region. Thus, we do not see
evidence for an ``overmerging problem'' as proposed by Leon et al.\
(\cite{lbv}).  

Our findings are clearly in favour of the formation scenario proposed by
Fujimoto \& Kumai (\cite{fk}) who suggested that the components of a binary
cluster formed together, and thus should be coeval or at least have a
small age difference compatible with cluster formation time scales. 

\begin{acknowledgements}
The authors acknowledge J\"org Sanner, Klaas~S.\ de\,Boer and Lindsay King for
carefully reading the manuscript of this publication. 
AD thanks Daniel Harbeck for the introduction to the methods of KMM. 
This paper has made use of the OGLE database of LMC star clusters. We
are grateful to the OGLE collaboration for making their data publicly
available.
This research has
made use of NASA's Astrophysics Data System Bibliographic Services, the CDS
data archive in Strasbourg, France.
\end{acknowledgements}

\newpage

\setlength{\tabcolsep}{0.1cm}
\begin{table*}[h!]
\caption{\label{groupscatalogue}{Catalogue of all binary and multiple cluster
    candidates found in the entire LMC area. Identifiers and remarks,
    coordinates, object type, maximum and minimum diameter ($D_{\rm
    max}$ and $D_{\rm min}$) and the position angle (P.A.) are taken
    from BSDO. For the acronyms of the objects see BSDO, their Table
    1. The acronym used in the OGLE catalogue of star clusters in the
    LMC (e.g., LMC0012, Pietrzy\'{n}ski \& Udalski \cite{pu_lmc}) is
    also given. The $9^{\rm th}$ column gives the separations ($d$) in
    pc found in the corresponding group, assuming a distance modulus
    of 18.5 mag. The last column gives the ages available in the
    literature, the notes are explained at the end of this paper. In
    some cases, only an age for the association of which the cluster
    appears to be part is found. If so, a corresponding remark is
    given in brackets. Only the first 21 groups are listed, the
    complete table can be found at CDS, Strasbourg}}  
\centerline{  
\begin{tabular}{rlcccccrrl}
\hline
no. & identifiers \& remarks & $\alpha$  & $\delta$ & type & $D_{\rm max}$ & $D_{\rm min}$ & P.A. & \multicolumn{1}{c}{$d$} & age   \\
    &  & [$^{\rm h} ~^{\rm m} ~^{\rm s}$] & [$^{\circ} ~\arcmin ~\arcsec$] &  & [$\arcmin$] & [$\arcmin$] & [$^{\circ}$] & \multicolumn{1}{c}{[pc]} & [Myr] \\
\hline
  1 &\raggedright{SL23, LW36, HS24, BRHT23b, KMHK50}   & 4 43 38 & --69 42 44 & CA & 1.10 & 0.95 & 100 &      9.1  & -- \\\nopagebreak
  1 & SL23A, BRHT23a, KMHK52                           & 4 43 43 & --69 43 11 & CA & 0.85 & 0.70 &  60 &      9.1  & -- \\
\hline                                                                                                                       
  2 & BSDL8                                            & 4 43 59 & --68 45 22 & AC & 0.65 & 0.50 & 150 &      9.0  & -- \\\nopagebreak
  2 & BSDL9                                            & 4 43 59 & --68 45 59 & AC & 0.45 & 0.35 & 100 &      9.0  & -- \\\nopagebreak
  2 & LW39, KMHK54                                     & 4 43 59 & --68 46 43 & CA & 0.95 & 0.85 &  60 &     19.7  & --  \\\nopagebreak
  2 & LW41, KMHK59                                     & 4 44 11 & --68 44 57 &  C & 0.90 & 0.90 &  -- &     17.0  & -- \\
\hline                                                                                                                       
  3 & BSDL10                                           & 4 44 05 & --69 52 50 &  C & 0.35 & 0.30 & 110 &     19.9  & --   \\\nopagebreak
  3 & SL24, LW38, KMHK55                               & 4 43 50 & --69 52 23 &  C & 1.10 & 1.10 &  -- &     19.9  & -- \\
\hline                                                                                                                       
  4 & BRHT59a, KMHK61 (in BSDL7)          & 4 44 13 & --71 22 01 &  C & 0.90 & 0.65 & 140 &     10.3  & $>$600 (x) \\\nopagebreak
  4 & LW43, BRHT59b, KMHK62 (in BSDL7)  & 4 44 16 & --71 22 41 &  C & 0.90 & 0.90 &  -- &     10.3  & $>$600 (x) \\
\hline                                                                                                                       
  5 & BSDL14                                           & 4 44 59 & --70 18 19 & CA & 0.50 & 0.40 &  40 &     16.3  & -- \\\nopagebreak
  5 & BSDL15                                           & 4 44 59 & --70 19 26 & CA & 0.55 & 0.50 & 160 &     16.3  & -- \\
\hline                                                                                                                       
  6 & LW56e, KMHK83e                                   & 4 45 54 & --72 21 08 &  C & 0.50 & 0.50 &  -- &      2.4  & -- \\\nopagebreak
  6 & LW56w, KMHK83w                                   & 4 45 52 & --72 21 04 &  C & 0.60 & 0.60 &  -- &      2.4  & -- \\
\hline                                                                                                                       
  7 & BSDL25                                           & 4 46 24 & --72 33 28 & AC & 0.85 & 0.75 & 140 &      9.1  & -- \\\nopagebreak
  7 & SL33, LW59, KMHK91                               & 4 46 25 & --72 34 05 &  C & 1.10 & 1.10 &  -- &      9.1  & -- \\
\hline                                                                                                                       
  8 & BSDL55 (in BSDL56)                               & 4 49 25 & --69 27 55 & AC & 0.70 & 0.55 &  70 &     11.6  & -- \\\nopagebreak
  8 & HS34 (in BSDL56)                                 & 4 49 31 & --69 28 31 & AC & 0.50 & 0.40 &  10 &     11.6  & -- \\
\hline                                                                                                                       
  9 & KMHK136                                          & 4 50 12 & --68 59 49 & AC & 1.00 & 0.85 &  80 &      3.0  & 10--30 (e) \\\nopagebreak
  9 & SL49 in KMHK136                                  & 4 50 10 & --68 59 55 &  C & 0.70 & 0.60 & 170 &      3.0  & 10--30 (e) \\
\hline                                                                                                                       
 10 & BSDL104 (in NGC1712)                             & 4 51 10 & --69 23 42 & CN & 0.65 & 0.55 & 110 &     13.9  & 0--10 (NGC1712) (e),  20 (y) \\\nopagebreak
 10 & BSDL96 (in NGC1712)                              & 4 51 01 & --69 23 10 &  C & 0.90 & 0.75 &  70 &     13.9  & 0--10 (NGC1712) (e)  \\
\hline                                                                                                                       
 11 & LW75, SL59w, KMHK152, BRHT24a       & 4 50 14 & --73 38 47 &  C & 1.20 & 1.10 &   0 &     11.4  & 1200 (s),  $>$600 (x) \\\nopagebreak
 11 & LW76, SL59e, KMHK157, BRHT24b       & 4 50 25 & --73 38 53 &  C & 1.20 & 1.10 &  20 &     11.4  & 1200 (s),  $>$600 (x) \\
\hline                                                                                                                       
 12 & BSDL100 (in BSDL101)                             & 4 50 58 & --70 00 30 & AC & 0.50 & 0.35 &  20 &     13.4  & -- \\\nopagebreak
 12 & BSDL103 (in BSDL101)                             & 4 51 03 & --70 00 43 & CA & 0.50 & 0.35 &  60 &      7.0  & -- \\\nopagebreak
 12 & KMHK156 (in SGshell LMC7)                        & 4 51 00 & --70 01 24 & CA & 0.90 & 0.80 & 120 &     13.4  & -- \\
\hline                                                                                                                       
 13 & KMHK164 (in BSDL110)                             & 4 51 23 & --69 35 04 &  C & 0.50 & 0.45 & 130 &     16.0  & -- \\\nopagebreak
 13 & KMHK166 (in BSDL110)                             & 4 51 32 & --69 34 18 & AC & 0.55 & 0.55 &  -- &     16.0  & -- \\
\hline                                                                                                                       
 14 & BSDL120 (in LMC N79A)                            & 4 51 47 & --69 23 14 & NC & 0.85 & 0.70 &  10 &      7.2  & -- \\\nopagebreak
 14 & BSDL124 (in BRHT1a)                              & 4 51 51 & --69 24 02 & NC & 0.50 & 0.40 & 110 &     12.7  & 0--10 (e),  25 (r) \\\nopagebreak
 14 & BSDL126 (in LMC N79A)                            & 4 51 53 & --69 23 26 & NC & 0.65 & 0.50 & 140 &      8.2  & -- \\\nopagebreak
 14 & IC2111, ESO56EN13, BRHT1b  & 4 51 51 & --69 23 35 & NC & 0.65 & 0.55 & 130 &      7.2  & 2--3 (F),  3.7--4.3 (i) \\\nopagebreak
  & (in LMC\,N79A) & & & & & & & & \\\nopagebreak
 14 & KMHK171 (in BRHT1a)                              & 4 51 53 & --69 24 24 & NC & 0.90 & 0.90 &  -- &     18.7  & 0--10 (e),  25 (r) \\\nopagebreak
 14 & LMC--N79B (in NGC1722=BRHT1a)       & 4 52 00 & --69 23 43 & NC & 0.40 & 0.35 & 140 &     18.1  & 0--10 (e),  25 (r) \\
\hline                                                                                                                       
 15 & BSDL129                                          & 4 51 56 & --70 23 52 & CA & 0.40 & 0.35 &  70 &      7.4  & -- \\\nopagebreak
 15 & SL66, KMHK180                                    & 4 51 55 & --70 23 22 &  C & 1.20 & 1.10 & 130 &      7.4  & 2000 -- 5000 (e) \\
\hline                                                                                                                       
 16 & H88--11, H80F1--10                               & 4 52 20 & --68 59 32 & AC & 0.50 & 0.35 &  30 &     16.8  & -- \\\nopagebreak
 16 & H88--7, H80F1--8                                 & 4 52 12 & --69 00 26 & AC & 0.55 & 0.40 &  80 &     16.8  & -- \\
\hline                                                                                                                       
 17 & BSDL155 (in LMC DEM13)                           & 4 53 13 & --68 01 48 & AC & 0.50 & 0.40 & 130 &     19.9  & -- \\\nopagebreak
 17 & HDE268680 (in NGC1736)                           & 4 53 03 & --68 03 06 & NC & 0.95 & 0.80 & 110 &      6.8  & 0--10 (e) \\\nopagebreak
 17 & LMC--S6 (in NGC1736)                             & 4 53 08 & --68 03 05 & NC & 0.35 & 0.35 &  -- &      6.8  & 0--10 (e) \\
\hline                                                                                                                       
 18 & BSDL157 (in SGshell LMC7)                        & 4 53 00 & --69 38 42 & AC & 0.50 & 0.45 &  60 &     18.7  & -- \\\nopagebreak
 18 & KMHK207 (in SGshell LMC7)                        & 4 53 00 & --69 37 25 &  C & 0.75 & 0.75 &  -- &     18.7  & -- \\
\hline                                                                                                                       
 19 & BSDL158                                          & 4 53 09 & --68 38 34 & AC & 0.50 & 0.50 &  -- &      7.1  & -- \\\nopagebreak
 19 &NGC1732, SL77, ESO56SC17, KMHK209   & 4 53 11 & --68 39 01 &  C & 1.10 & 1.00 &  50 &      7.1  & 30 -- 70 (e) \\
\hline                                                                                                                       
 20 & KMHK212 (in NGC1731)                             & 4 53 35 & --66 55 25 &  C & 0.75 & 0.60 & 170 &      8.6  & $<$4 (NGC1731) (G) \\\nopagebreak
 20 & SL82, KMHK211 (in NGC1731)          & 4 53 29 & --66 55 28 & AC & 0.85 & 0.60 & 100 &      8.6  & $<$4 (NGC1731) (G) \\
\hline                                                                                                                       
 21 & BSDL162                                          & 4 53 24 & --67 53 00 & AC & 0.55 & 0.55 &  -- &     19.1  & -- \\\nopagebreak
 21 & HS56, KMHK218                                    & 4 53 36 & --67 52 20 & CA & 0.75 & 0.70 &  50 &     19.1  & -- \\
\hline
\end{tabular}}
\end{table*} 

Notes to Table~\ref{groupscatalogue}:

\parindent0cm
(a): Banks et al.\ (\cite{banks}): $BV$ CMD and isochrone fitting\\
(b): Barbaro \& Olivi (\cite{barbaro}): UV spectra of the clusters and comparison with models\\
(c): Bhatia (\cite{bhatia92}): integrated $BVR$ photometry\\
(d): Bhatia \& Piotto (\cite{bp}): $BV$ CMD and isochrone fitting\\
(e): Bica et al.\ (\cite{bcdsp}): integrated $UV$ photometry\\
(f): Caloi \& Cassatella (\cite{cc}): IUE spectra, CMD and evolutionary tracks \\
(g): Cassatella et al.\ (\cite{cassatella}): integrated UV colours \\
(h): Chiosi et al.\ (\cite{chiosi}): integrated $UBV$ colours, synthetic HR diagrams, turn-off ages from Chiosi et al.\ (\cite{chiosi86})\\
(i): Copetti et al.\ (\cite{copetti}): age estimates from [O\,III]/H$\beta$ for H\,II regions\\
(j): de\,Oliveira et al.\ (\cite{dodb}): ages from SWB types deduced from $UBV$ colours in Alcaino (\cite{alcaino})\\
(k): Dieball \& Grebel (\cite{dg}): CMD and isochrone fitting \\
(l): Dieball et al.\ (\cite{dgt}): CMD and isochrone fitting\\
(m): Dieball \& Grebel (\cite{dg2000}): CMD and isochrone fitting\\
(n): Dirsch et al.\ (\cite{dirsch}): Str\"omgren CCD photometry and isochrone fitting\\
(o): Elson \& Fall (\cite{elson88}): integrated $UBV$ colours\\
(p): Elson (\cite{elson91}): CMDs and isochrone fitting\\
(q): Fischer  et al.\ (\cite{fwm}): $BV$ CMD and isochrone fitting \\
(r): Fujimoto \& Kumai (\cite{fk}): ages from $U-B$, $B-V$ TCDs and synthetic evolutionary models\\ 
(s): Geisler et al.\ (\cite{geisler97}): $\delta T$ magnitude difference between turn-off and giant branch clump\\
(t): Gilmozzi et al.\ (\cite{gilmozzi}): CMD and isochrone fitting\\
(u): Girardi et al.\ (\cite{girardi}): integrated $UBV$ colours\\
(v): Hilker et al.\ (\cite{hrs}): Str\"omgren CCD photometry and isochrone fitting\\
(w): Kontizas et al.\ (\cite{kkd}): HR diagram and isochrone fitting\\
(x): Kontizas et al.\ (\cite{kkm}): integrated IUE spectra and stellar content\\
(y): Kubiak (\cite{kubiak}): CMD and isochrone fitting\\
(z): Laval et al.\ (\cite{laval}): H$\alpha$ observations, kinematical data\\
(A): Laval et al.\ (\cite{lrb}): H$\alpha$ observations, kinematical data\\
(B): Laval et al.\ (\cite{lgg}): $VBLUW$ colours, isochrone and comparison with cluster of known age (NGC\,6231)\\
(C): Lee (\cite{lee92}): $UBVI$ photometry\\
(D): Meurer et al.\ (\cite{meurer}): UV colours as age indicator\\
(E): Oliva \& Origlia (\cite{oliva}): IR spectra, age from Elson \& Fall (\cite{elson88})\\
(F): Santos et al.\ (\cite{sbc}): integrated blue-violet spectral evolution, Table 6 \\
(G): Santos et al.\ (\cite{sbc}): from $U-B$ calibration, Table 1 \\
(H): Shull (\cite{shull}): age from kinematic considerations\\
(I): Tarrab (\cite{tarrab}): ages from H$\beta$ equivalent width ($W_{\rm H}\beta$)\\
(J): Testor et al.\ (\cite{testor}): HR diagram\\
(K): Vallenari et al.\ (\cite{vafcom}): CMD and isochrone fitting\\
(L): Vallenari et al.\ (\cite{vbc}): CMD and isochrone fitting\\
(M): Will et al.\ (\cite{wvfs}): CCD photometry \\
(N): Piertzy\'{n}ski \& Udalsky (\cite{pu_lmc}): $BVI$ CCD data and
isochrone fitting 

\end{document}